\documentclass[twocolumn,twocolappendix]{aastex63}

\usepackage{natbib,aas_macros,amsmath}
\usepackage{multirow,color}
\usepackage{natbib}

\def\farcs{%
 \mbox{%
  \kern  0.13ex.%
  \kern -0.95ex\arcsec%
  \kern -0.1ex%
 }%
}%

\def\tcb{\textcolor{black}}

\newcommand{\oiii}{[O\,{\sc iii}]\,$\lambda$}
\newcommand{\oiiib}{O\,{\sc iii}]\,$\lambda\lambda$}

\newcommand{\oii}{[O\,{\sc ii}]\,$\lambda\lambda$}

\newcommand{\ciii}{C\,{\sc iii}]\,$\lambda\lambda$}
\newcommand{\civ}{C\,{\sc iv}\,$\lambda$}
\newcommand{\nii}{[N\,{\sc ii}]}
\newcommand{\neiii}{[Ne\,{\sc iii}]\,$\lambda$}
\newcommand{\nev}{[Ne\,{\sc v}]\,$\lambda$}
\newcommand{\niv}{N\,{\sc iv}]\,$\lambda$}
\newcommand{\heii}{He\,{\sc ii}\,$\lambda$}

\newcommand{\hst}{{\it HST}}
\newcommand{\jwst}{{\it JWST}}
\newcommand{\prospector}{\texttt{Prospector}}
\newcommand{\eazy}{\texttt{EAZY}}

\submitjournal{ApJ }
\shorttitle{\jwst/NIRSpec Census at $z=8.5-13.1$ in UNCOVER}
\shortauthors{Fujimoto et al.}

\begin{document}

\title{
UNCOVER: A NIRSpec Census of Lensed Galaxies at \boldmath $z=8.50$--13.08 \\ 
Probing a High AGN Fraction and Ionized Bubbles in the Shadow
}

\correspondingauthor{Seiji Fujimoto}
\email{fujimoto@utexas.edu}
\author[0000-0001-7201-5066]{Seiji Fujimoto}\altaffiliation{Hubble Fellow}
\affiliation{
Department of Astronomy, The University of Texas at Austin, Austin, TX 78712, USA
}

\author[0000-0001-9269-5046]{Bingjie Wang}
\affiliation{Department of Astronomy \& Astrophysics, The Pennsylvania State University, University Park, PA 16802, USA}
\affiliation{Institute for Computational \& Data Sciences, The Pennsylvania State University, University Park, PA 16802, USA}
\affiliation{Institute for Gravitation and the Cosmos, The Pennsylvania State University, University Park, PA 16802, USA}

\author[0000-0003-1614-196X]{John R. Weaver}
\affiliation{Department of Astronomy, University of Massachusetts, Amherst, MA 01003, USA}

\author[0000-0002-5588-9156]{Vasily Kokorev}
\affiliation{Kapteyn Astronomical Institute, University of Groningen, 9700 AV Groningen, The Netherlands}

\author[0000-0002-7570-0824]{Hakim Atek}
\affiliation{Institut d'Astrophysique de Paris, CNRS, Sorbonne Universit\'e, 98bis Boulevard Arago, 75014, Paris, France}

\author[0000-0001-5063-8254]{Rachel Bezanson}
\affiliation{Department of Physics and Astronomy and PITT PACC, University of Pittsburgh, Pittsburgh, PA 15260, USA}

\author[0000-0002-2057-5376]{Ivo Labbe}
\affiliation{Centre for Astrophysics and Supercomputing, Swinburne University of Technology, Melbourne, VIC 3122, Australia}

\author[0000-0003-2680-005X]{Gabriel Brammer}
\affiliation{Cosmic Dawn Center (DAWN), Niels Bohr Institute, University of Copenhagen, Jagtvej 128, K{\o}benhavn N, DK-2200, Denmark}

\author[0000-0002-5612-3427]{Jenny E. Greene}
\affiliation{Department of Astrophysical Sciences, Princeton University, 4 Ivy Lane, Princeton, NJ 08544}

\author[0009-0009-9795-6167]{Iryna Chemerynska}
\affiliation{Institut d'Astrophysique de Paris, CNRS, Sorbonne Universit\'e, 98bis Boulevard Arago, 75014, Paris, France}

\author[0000-0001-8460-1564]{Pratika Dayal}
\affiliation{Kapteyn Astronomical Institute, University of Groningen, 9700 AV Groningen, The Netherlands}

\author[0000-0002-2380-9801]{Anna de Graaff}
\affiliation{Max-Planck-Institut f{\"u}r Astronomie, K{\"o}nigstuhl 17, D-69117, Heidelberg, Germany}

\author[0000-0001-6278-032X]{Lukas J. Furtak}
\affiliation{Physics Department, Ben-Gurion University of the Negev, P.O. Box 653, Be'er-Sheva 84105, Israel}

\author[0000-0001-5851-6649]{Pascal A. Oesch}
\affiliation{Department of Astronomy, University of Geneva, Chemin Pegasi 51, 1290 Versoix, Switzerland}
\affiliation{Cosmic Dawn Center (DAWN), Niels Bohr Institute, University of Copenhagen, Jagtvej 128, K{\o}benhavn N, DK-2200, Denmark}

\author[0000-0003-4075-7393]{David J. Setton}
\affiliation{Department of Physics and Astronomy and PITT PACC, University of Pittsburgh, Pittsburgh, PA 15260, USA}

\author[0000-0002-0108-4176]{Sedona H. Price}
\affiliation{Department of Physics and Astronomy and PITT PACC, University of Pittsburgh, Pittsburgh, PA 15260, USA}

\author[0000-0001-8367-6265]{Tim B. Miller}
\affiliation{Department of Astronomy, Yale University, New Haven, CT 06511, USA}
\affiliation{Center for Interdisciplinary Exploration and Research in Astrophysics (CIERA) and
Department of Physics and Astronomy, Northwestern University, 1800 Sherman Ave, Evanston IL 60201, USA}

\author[0000-0003-2919-7495]{Christina C. Williams}
\affiliation{NSF’s National Optical-Infrared Astronomy Research Laboratory, 950 N. Cherry Avenue, Tucson, AZ 85719, USA}
\affiliation{Steward Observatory, University of Arizona, 933 North Cherry Avenue, Tucson, AZ 85721, USA}

\author[0000-0001-7160-3632]{Katherine E. Whitaker}
\affiliation{Department of Astronomy, University of Massachusetts, Amherst, MA 01003, USA}

\author[0000-0002-0350-4488]{Adi Zitrin}
\affiliation{Physics Department, Ben-Gurion University of the Negev, P.O. Box 653, Be'er-Sheva 84105, Israel}

\author[0000-0002-7031-2865]{Sam E. Cutler}
\affiliation{Department of Astronomy, University of Massachusetts, Amherst, MA 01003, USA}

\author[0000-0001-6755-1315]{Joel Leja}
\affiliation{Department of Astronomy \& Astrophysics, The Pennsylvania State University, University Park, PA 16802, USA}
\affiliation{Institute for Computational \& Data Sciences, The Pennsylvania State University, University Park, PA 16802, USA}
\affiliation{Institute for Gravitation and the Cosmos, The Pennsylvania State University, University Park, PA 16802, USA}

\author[0000-0002-9651-5716]{Richard Pan}
\affiliation{Department of Physics and Astronomy, Tufts University, 574 Boston Ave., Medford, MA 02155, USA}

\author[0000-0001-7410-7669]{Dan Coe}
\affiliation{Space Telescope Science Institute (STScI), 3700 San Martin Drive, Baltimore, MD 21218, USA}
\affiliation{Association of Universities for Research in Astronomy (AURA), Inc. for the European Space Agency (ESA)}
\affiliation{Center for Astrophysical Sciences, Department of Physics and Astronomy, The Johns Hopkins University, 3400 N Charles St. Baltimore, MD 21218, USA}

\author[0000-0002-8282-9888]{Pieter van Dokkum}
\affiliation{Department of Astronomy, Yale University, New Haven, CT 06511, USA}

\author[0000-0002-1109-1919]{Robert Feldmann}
\affiliation{Institute for Computational Science, University of Zurich, Zurich, CH-8057, Switzerland}

\author[0000-0001-7440-8832]{Yoshinobu Fudamoto} 
\affiliation{Waseda Research Institute for Science and Engineering, Faculty of Science and Engineering, Waseda University, 3-4-1 Okubo, Shinjuku, Tokyo 169-8555, Japan}
\affiliation{National Astronomical Observatory of Japan, 2-21-1, Osawa, Mitaka, Tokyo, Japan}

\author[0000-0003-4700-663X]{Andy D. Goulding}
\affiliation{Department of Astrophysical Sciences, Princeton University, 4 Ivy Lane, Princeton, NJ 08544}

\author[0000-0002-3475-7648]{Gourav Khullar}
\affiliation{Department of Physics and Astronomy and PITT PACC, University of Pittsburgh, Pittsburgh, PA 15260, USA}

\author[0000-0001-9002-3502]{Danilo Marchesini}
\affiliation{Physics and Astronomy Department, Tufts University, 574 Boston Ave., Medford, MA 02155, USA}

\author[0000-0003-0695-4414]{Michael Maseda}
\affiliation{Department of Astronomy, University of Wisconsin-Madison, 475 N. Charter St., Madison, WI 53706 USA}

\author[0000-0003-2804-0648]{Themiya Nanayakkara}
\affiliation{Centre for Astrophysics and Supercomputing, Swinburne University of Technology, Melbourne, VIC 3122, Australia}

\author[0000-0002-7524-374X]{Erica J. Nelson}
\affiliation{Department for Astrophysical and Planetary Science, University of Colorado, Boulder, CO 80309, USA}

\author[0000-0001-8034-7802]{Renske Smit}
\affiliation{Astrophysics Research Institute, Liverpool John Moores University, 146 Brownlow Hill, Liverpool L3 5RF, UK}

\author[0000-0001-7768-5309]{Mauro Stefanon}
\affiliation{Departament d'Astronomia i Astrof\`isica, Universitat de Val\`encia, C. Dr. Moliner 50, E-46100 Burjassot, Val\`encia,  Spain}
\affiliation{Unidad Asociada CSIC "Grupo de Astrof\'isica Extragal\'actica y Cosmolog\'ia" (Instituto de F\'isica de Cantabria - Universitat de Val\`encia)}

\author[0000-0001-8928-4465]{Andrea Weibel}
\affiliation{Department of Astronomy, University of Geneva, Chemin Pegasi 51, 1290 Versoix, Switzerland}


\def\apj{ApJ}%
\def\apjl{ApJ}%
\def\apjs{ApJS}%

\def\rme{\rm e}
\def\rmstar{\rm star}
\def\rmFIR{\rm FIR}
\def\itHubble{\it Hubble}
\def\rmyr{\rm yr}

\begin{abstract}
We present \jwst\ NIRSpec prism spectroscopy of lensed galaxies at $z\gtrsim9$ found behind the massive galaxy cluster Abell~2744 in the UNCOVER Cycle~1 Treasury Program. We confirm the redshift via emission lines and/or the Ly$\alpha$ break for ten galaxies at $z=8.$50--13.08 down to $M_{\rm\,UV}=-17.3$. We achieve a 100\% confirmation rate for $z>9$ candidates reported in \cite{atek2023}. Using six sources with multiple line detections, we find that offsets in redshift estimates between the lines and the Ly$\alpha$ break alone can be $\pm0.2$, raising caution in designing future follow-up spectroscopy for the break-only sources \tcb{with ALMA}. 
With spec-$z$ confirmed sources in UNCOVER and literature, we derive lower limits on the rest-frame ultraviolet (UV) luminosity function (LF) at $z\simeq9$--12 and find that these lower limits agree with recent photometric measurements. We identify at least two unambiguous and several possible active galactic nucleus (AGN) systems based on X-ray, broad H$\beta$, high ionization lines, and excess in UVLF. This requires the AGN LFs at $z\simeq$9--10 to be comparable or even higher than the X-ray AGN LF estimated at $z\sim6$ and suggests a plausible cause of the high abundance of $z>9$ galaxies claimed in the recent photometric measurements is AGNs. One UV-luminous source is confirmed at the same redshift as a broad-line AGN at $z=8.50$ with a physical separation of 380~kpc in the source plane. These two sources show the emission blueward of Ly$\alpha$, indicating a giant ionized bubble enclosing them with a radius of $7.69\pm0.18$~pMpc. Our results imply that AGNs have a non-negligible contribution to cosmic reionization.
\end{abstract}



\section{Introduction}
\label{sec:intro} 

Studying early galaxies provides key clues to understanding fundamental cosmological questions such as dark matter assembly, the development of large-scale structure, the emergence of the first galaxies and black holes, and the processes that govern cosmic reionization and early galaxy formation and evolution \citep[e.g.,][]{dayal2018, inayoshi2020}. 
In the last decades, deep \textit{Hubble Space Telescope (HST)} surveys have succeeded in discovering thousands of galaxies in the Epoch of Reionization (EoR) $6\lesssim z\lesssim11$, providing valuable photometric insights for these galaxies, including an initial characterization of the stellar component, in terms of un-obscured star formation rates and sizes \citep[e.g.,][]{ellis2013, bouwens2015, finkelstein2015, oesch2016, bhatawdekar2019}. 

The advent of \jwst\ \citep{gardner2023} has led to significant progress in discovering and investigating galaxies at very early cosmic epochs. 
As demonstrated in the Early Release Observations (ERO; \citealt{pontoppidan2022}) and the Early Release Science programs (ERS; e.g., \citealt{treu2022, finkelstein2023}), 
dozens of high-redshift galaxy candidates have been identified at $z\simeq$ 9--17 towards both lensing clusters and blank fields \citep[e.g.,][]{adams2022, atek2022, atek2023, bouwens2022c, bradley2022, castellano2022, donnan2023, finkelstein2022b, finkelstein2023, harikane2023, labbe2022, morishita2022, naidu2022, yan2022, williams2023, austin2023, leung2023}.  
Their abundance at the bright-end ($M_{\rm UV} \lesssim -20$) exceeds nearly all theoretical predictions so far \citep[e.g.,][]{behroozi2015, dayal2017,   yung2019a, yung2020b, behroozi2019, behroozi2020, dave2019, wilkins2022a, wilkins2022b, kannan2022,  mason2022, mauerhofer2023}, 
suggesting several possibilities, including that star formation in early systems is dominated by a top-heavy initial mass function (IMF), complete lack of dust attenuation, 
stochastic star-formation, and/or the emergence of the active galactic nucleus (AGN) population \citep[e.g.,][]{harikane2023, finkelstein2023, pacucci2022, ferrara2022, boylan2022, lovell2022, menci2022, sun2023}. 

\jwst\ NIRSpec follow-up spectroscopy has been performed for several bright galaxy candidates at $z\simeq10$--17, including Director’s Discretionary Time (DDT). 
These follow-up observations confirm the source redshifts via emission lines and/or the unambiguous Lyman-$\alpha$ break feature at $z=9.5$--13.2 \citep{williams2022, roberts-borsani2022b, bunker2023, curtis-lake2023, arrabal-halo2023a, arrabal-haro2023b, hsiao2023, harikane2023b, bwang2023}. 
However, one remarkably UV-bright ($M_{\rm UV}\simeq-22$) galaxy candidate at $z\simeq16$ turns out to be $z=4.9$ with strong emission lines and red continuum that mimic the expected colors of more distant objects \citep{naidu2022b, zavala2023, fujimoto2022b, mackinney2023, arrabal-halo2023a}. 
Recent \jwst\ spectroscopic observations also find a numerous number of faint AGN populations at $z\simeq4$--7 \citep[e.g.,][]{kocevski2023, harikane2023c, matthee2023, maiolino2023c}, indicative of steeper faint ends in the quasar/AGN LFs than suggested in previous studies, and some studies suggest the identification of AGNs even at higher redshifts at $z\sim9-11$ \citep{larson2023, goulding2023, maiolino2023b}. 
These results indicate the critical importance of spectroscopy in order to foremost confirm the high-redshift nature of galaxy candidates and consequently verify earlier (photometric) claims of a high abundance of $z\gtrsim9$ galaxy candidates in order to investigate its origins.

In this paper, we present \jwst\ NIRSpec prism follow-up observations of $z\gtrsim9$ galaxy candidates, including an X-ray luminous supermassive black hole \citep{goulding2023} and two $z>12$ galaxies \citep{bwang2023}, all identified in the Cycle~1 Treasury program of Ultradeep NIRSpec and NIRCam Observations before the Epoch of Reionization survey (UNCOVER; \#2561, PIs: I.~Labbe \& R.~Bezanson; \citealt{bezanson2022}). 
This is the most extensive follow-up program with NIRSpec prism in Cycle~1 for \jwst-selected galaxy candidates towards a massive lensing cluster, providing a unique opportunity for a spectroscopic study for a large sample in the early universe over a wide UV luminosity range. 
Following the recent successful spectroscopic confirmation of the high redshift galaxies with \jwst, this enlarges the spectroscopic sample at $z\gtrsim9$ for faint sources ($M_{\rm UV}>-19$) by a factor of $\sim3$ and further allows detailed investigations into the UV luminosity function (LF) shape and the characterization of the high-redshift galaxy population newly identified with \jwst. 
In Section 2, we briefly describe observations and data processing.
Section 3 outlines our methods and results for the redshift measurements. 
In Section 4, we present our UV LF measurements at $z\simeq9$--12 and a potential high abundance of AGN.  
In Section 5, we report a discovery of a giant ionized bubble at $z=8.5$ and discuss the contribution of AGN to forming it. 
We summarize this study in Section 6.
Throughout this paper, we assume a flat universe with
$\Omega_{\rm m} = 0.3$,
$\Omega_\Lambda = 0.7$,
$\sigma_8 = 0.8$,
and $H_0 = 70$ km s$^{-1}$ Mpc$^{-1}$ \citep{hinshaw2013}.
All magnitudes are expressed in the AB system \citep{oke1983}. 
\tcb{The significance of X-ray detection in UHZ1 \citep{goulding2023} increases from follow-up \textit{Chandra} observations (Bogdan et al. in prep.), and we regard UHZ1 as the X-ray AGN throughout the paper.}

\section{Observations and Data processing} 
\label{sec:data}

The \jwst/NIRCam \citep{rieke2003, rieke2005, beichman2012, reike2023} and NIRSpec \citep{jakobsen2022} data employed in this paper were taken as a part of the UNCOVER survey \citep{bezanson2022}. While the complete descriptions of the NIRCam observations and NIRSpec observations are presented in \cite{weaver2023} and \cite{price2024}, respectively, we briefly describe the reduction employed in this paper in the following subsection. 

\subsection{UNCOVER Survey}
\label{sec:survey}

Abell 2744 (A2744) is among the most extensively studied massive galaxy clusters at $z=0.308$ and serves as the focal point of the UNCOVER survey. A2744 has been subjected to detailed observations using the \textit{Hubble Space Telescope} (\hst) as one of the clusters observed in the Hubble Frontier Field survey (HFF; \citealt{lotz2017}). 
The sky area of A2744 has a low infrared background, and its high magnification areas match well with the NIRCam field of view. Multiple \jwst\ Cycle~1 and 2 observations, including GTO, ERS, GO, and DDT programs, have been conducted and further planned towards this cluster. 
Specifically, a GO treasury program under the \jwst\ Cycle 1 -- UNCOVER (\#2561; PIs I. Labbe \& R. Bezanson) is designed to acquire in-depth NIRCam and NIRSpec observations over an extended $4'\times6'$ area \citep{bezanson2022}, enveloping the area with magnifications of $\mu\geq2$ encompassing the primary cluster observed in HFF and two supplementary cluster cores in the northern and north-west regions \citep{furtak2023a}. 
UNCOVER consists of two parts: 1) a deep NIRCam pre-imaging mosaic in 7 filters for $\sim4-6$ hours per band taken in November 2022, and 2) a $\sim24$-hour NIRSpec prism low-resolution follow-up of NIRCam-detected high-redshift galaxies in July-August 2023. 

\begin{figure}
\begin{center}
\includegraphics[trim=0cm 0cm 0cm 0cm, clip, angle=0,width=0.48\textwidth]{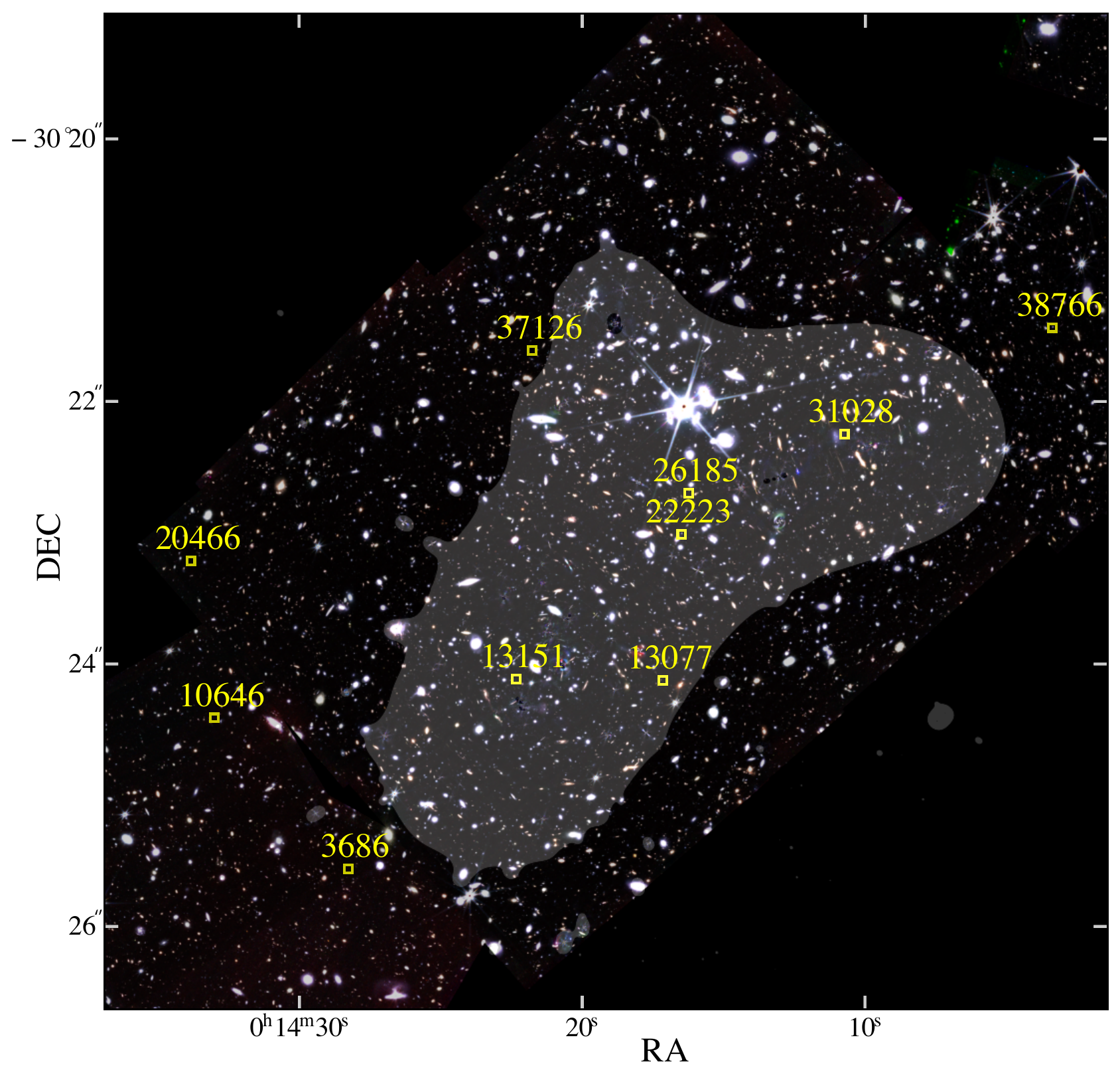}
\end{center}
\vspace{-0.6cm}
 \caption{
NIRCam RGB (R: F444W, G: F356W, B: F277W) map of A2744 taken in UNCOVER \citep{bezanson2022}. 
The white-shaded region indicates the highly magnified area with magnifications of $\geq2$ \citep{furtak2023a}.  
The yellow squares show the positions of the 10 sources that are spectroscopically confirmed at $z\geq8.5$ in UNCOVER. 
\label{fig:entire}}
\end{figure}

\begin{figure*}
\begin{center}
\includegraphics[trim=0cm 0cm 0cm 0cm, clip, angle=0,width=1.\textwidth]{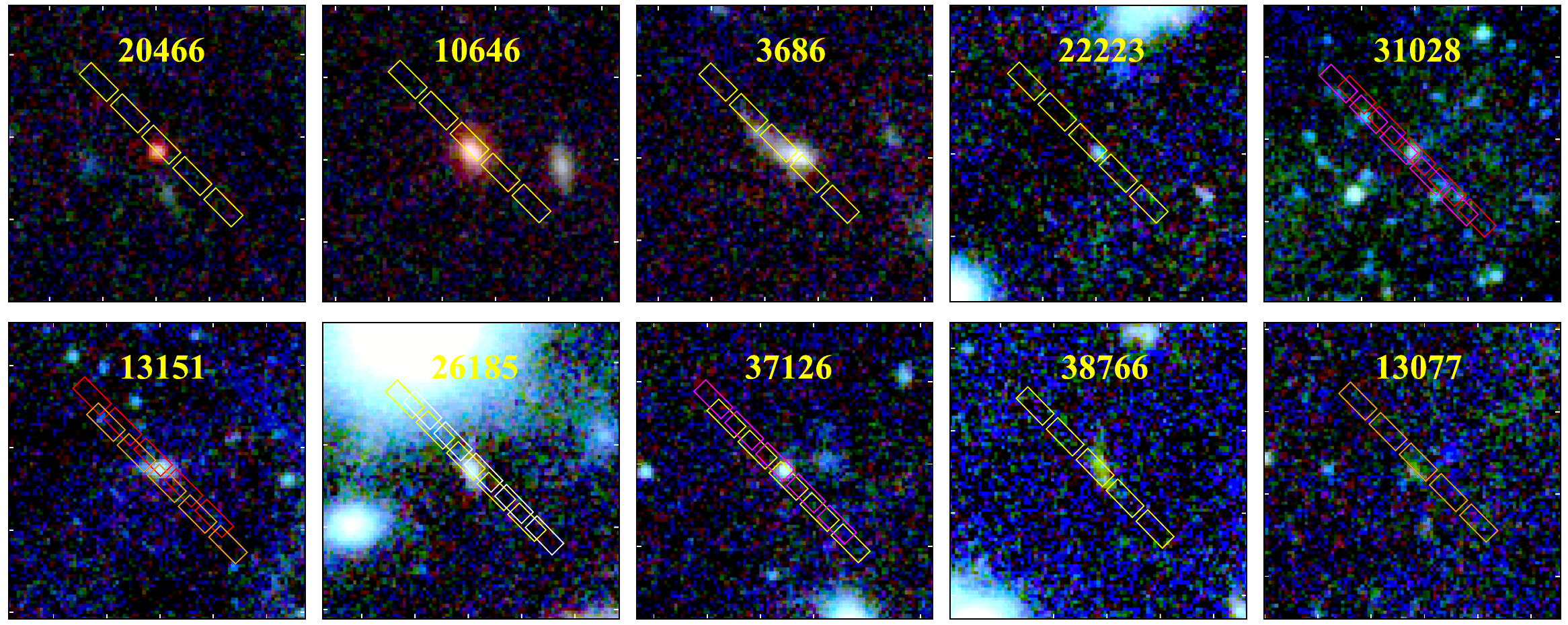}
\end{center}
\vspace{-0.4cm}
 \caption{
Zoom-in $3\farcs6\times3\farcs6$ NIRCam RGB (R: F444W, G: F356W, B: F277W) cutouts of the 10 sources whose spec-$z$ are successfully confirmed.  
The rectangles show the shutter configurations, where the standard three-shutter slitlets and a three-point nodding were adopted, and thus the five shutter positions are presented.  
The white, magenta, yellow, orange, red, and cyan rectangles represent our MSA observations of MSA-2, 3, 4, 5, 6, and 7, respectively, where MSA-5, 6, and 7 overlap nearly entirely. 
\label{fig:cutout}}
\end{figure*}

\subsection{NIRCam data \& Target selection}
\label{sec:nircam}

The galaxies discussed in this paper are selected from the UNCOVER NIRCam data taken in November 2022 \citep{bezanson2022}. 
The \cite{weaver2023} photometric catalog includes the measurements over the full NIRCam wavelength range in the F115W, F150W, F200W, F277W, F356W, F410M, and F444W filters, which have exposures of 3.7--6.0 hours per filter, as well as existing \hst\ Advanced Camera for Surveys (ACS) and WFC3 F606W, F814W, F105W, F125W, F140W, and F160W filters. 
The galaxy candidates at $z\gtrsim9$ are selected based on photometric redshift $z_{\rm photo}$ estimates from several SED analyses led by UNCOVER team members (e.g., \citealt{atek2023}; \citealt{bwang2024a}), 
while we also add several sources by visually checking the NIRCam SEDs and image cutouts in a less conservative sample selected based on photometric redshifts inferred from \eazy\ \citep{brammer2008} and \prospector\ \citep{johnson2021,bwang2023a} in order not to miss possible candidates. 
Together with other exciting high-redshift source candidates (e.g., faint AGNs, quiescent galaxies, strongly magnified and multiply imaged sources), the NIRSpec Multiobject Spectroscopy (MOS) configurations with the multi-shutter array (MSA) were designed to maximize the number of observed exciting candidates. 
We used seven MSA masks in our observations, referred to as MSA-1 through MSA-7. 
In this paper, we present ten sources whose redshifts are successfully confirmed at $z\geq8.5$ among a total of 680 MOS targets (Section~\ref{sec:measure}). 
In Figure~\ref{fig:entire}, we show the distribution of the ten galaxies in A2744, and Figure~\ref{fig:cutout} presents their NIRCam RGB color images with their MSA shutter configurations. 

\subsection{NIRSpec Data processing}
\label{sec:nirspec}

The data were reduced using the STScI \jwst\ pipeline for Level 1 data products and processed with \texttt{msaexp} \tcb{version v0.6.10}\footnote{\url{https://github.com/gbrammer/msaexp}} that is built on custom routines for further corrections in addition to the STScI pipeline routines to generate Level 2 and 3 products. 
The full descriptions of the NIRSpec data reduction will be presented in S.~Price et al., in preparation (see also \citealt{goulding2023, bwang2023}). 
Briefly, the data reduction was processed from the raw data files by using the Detector1Pipeline routine in combination with the latest batch of reference files (\texttt{jwst\_1100.pmap}) to correct detector-level artifacts and to convert the data into count rate images.  We then leverage custom pre-processing procedures from \texttt{msaexp} to correct for 1/$f$ noise, remove ``snowballs'' and bias on a per exposure basis before executing several STScI routines from the Spec2Pipeline to generate the final 2D cutout images. The AssignWcs, Extract2dStep, FlatFieldStep, and PhotomStep routines are utilized to perform WCS registration, flat fielding, and flux calibration. 
The PathLossStep accounting for the slitloss correction is turned off at this stage of the reduction process. Instead, we perform slitloss corrections by 
\tcb{fitting for a polynomial calibration vector of order 2 (after applying a wavelength-independent calibration to scale the normalization of the spectrum to the photometry)}. 
We undertake background subtraction locally, employing a three-shutter nod pattern before mapping the resultant images onto a uniform grid. 
From that point, we extract the spectra optimally through an inverse-variance weighted kernel derived by collapsing the 2D spectrum along the dispersion axis and fitting the ensuing signal along the spatial axis with a Gaussian profile.  
In some cases (ID31028, ID13151, ID13077), this Gaussian fitting is challenging due to the low SNR of the trace or the contamination of nearby sources in the shutters, and we manually set the Gaussian profile for the extraction. In Figure~\ref{fig:spectrum}, we show the 2$\sigma$ range of the Gaussian profile used to extract the 2D spectrum. 

\section{Spectroscopic Redshift}

\subsection{Measurements \& Results}
\label{sec:measure}

\begin{figure*}
\begin{center}
\includegraphics[trim=0cm 0cm 0cm 0cm, clip, angle=0,width=1.\textwidth]{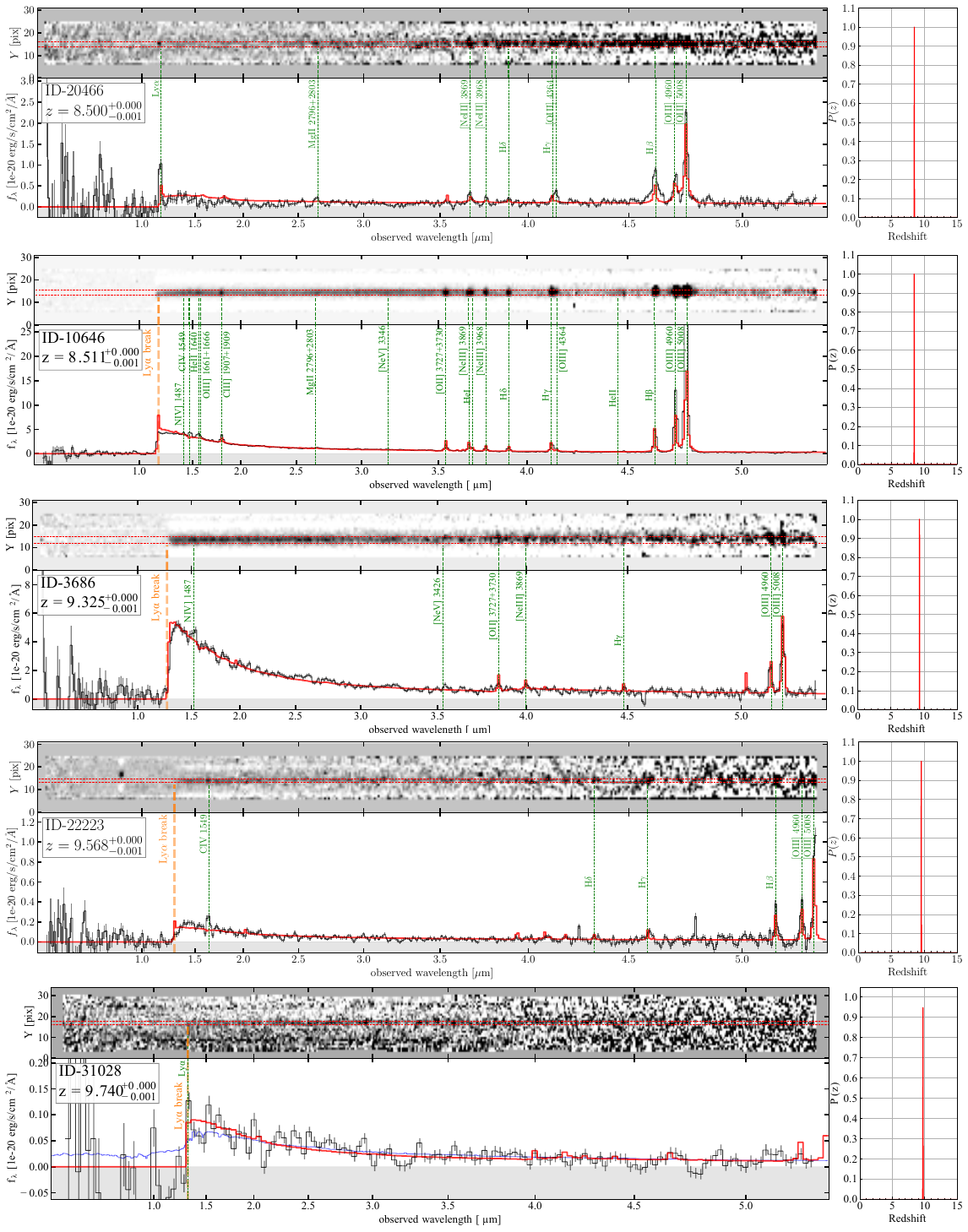}
\end{center}
\vspace{-0.8cm}
 \caption{
2D and 1D prism spectra for the ten sources whose spec-$z$ are successfully confirmed $z\geq8.5$ in this study. 
The red horizontal lines indicate the 2$\sigma$ range of the Gaussian, which is used for extracting the 1D spectrum shown in the bottom panel. 
The orange and green vertical lines denote wavelengths of the Ly$\alpha$ break and the faint emission lines detected at SNR $\geq$ 2.5, respectively. 
The right panel shows the likelihood of the source redshift $P(z)$ estimated from the \texttt{eazy} template fitting to the prism spectrum, and the best-fit SED (forced at $z<6$) is presented with the red (blue) curve overlaid on the 1D spectrum. 
\label{fig:spectrum}}
\end{figure*}

\begin{figure*}
\begin{center}
\includegraphics[trim=0cm 0cm 0cm 0cm, clip, angle=0,width=1.0\textwidth]{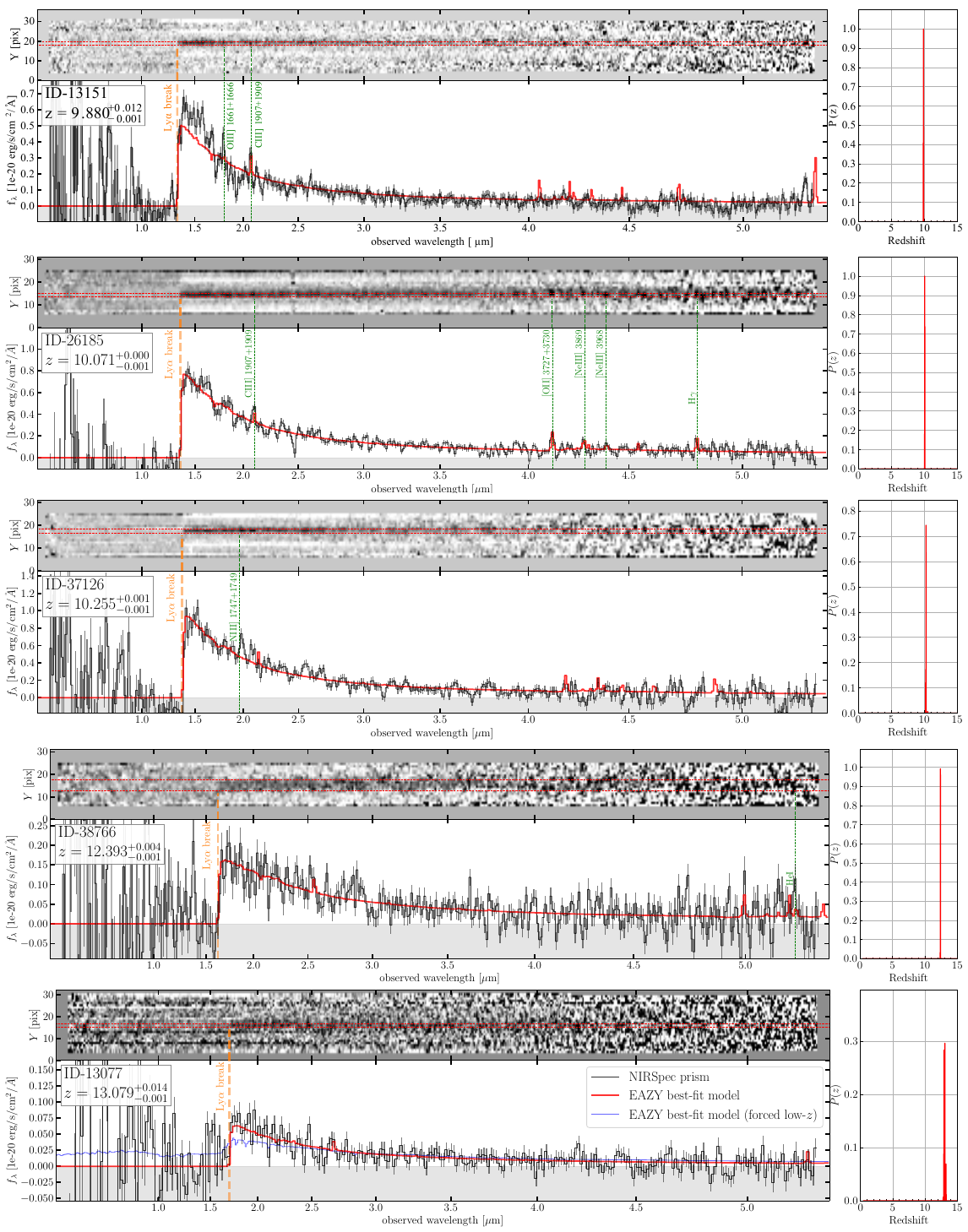}
{\bf Figure \ref{fig:spectrum}} (continued)
\end{center}
\end{figure*}

\begin{figure*}
\begin{center}
\includegraphics[trim=0cm 0cm 0cm 0cm, clip, angle=0,width=0.9\textwidth]{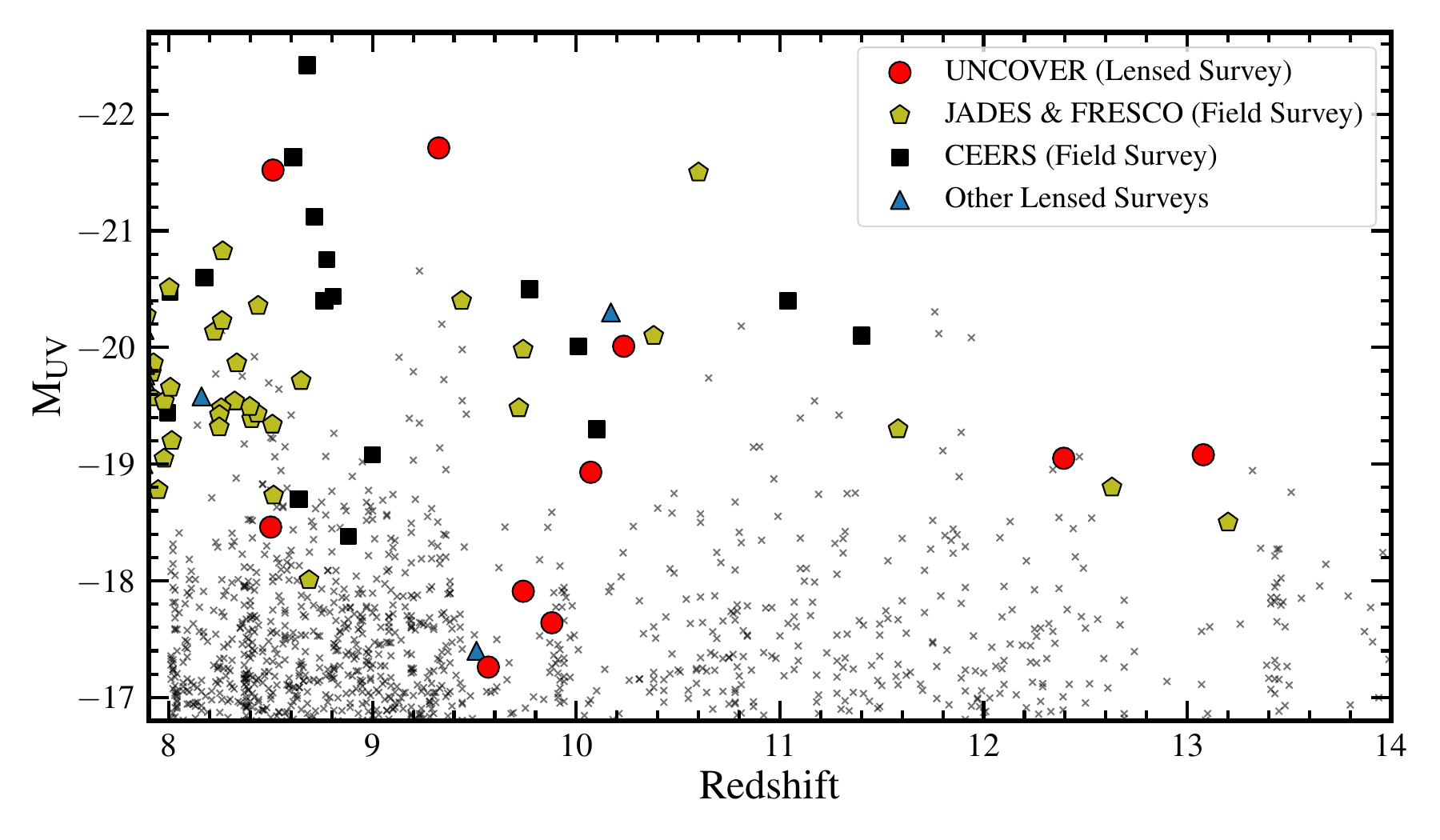}
\end{center}
\vspace{-0.6cm}
 \caption{
Redshift vs $M_{\rm UV}$. 
The magnification correction is applied for the lensed sources. 
The red circles represent our spec-$z$ confirmed sources in UNCOVER at $z\geq8.5$, efficiently increasing the sample at a faint ($M_{\rm UV}\gtrsim-19$) and high-redshift ($z\gtrsim9.5$) regime. 
The gold pentagons and the black squares show recent \jwst\ spectroscopic results from other field surveys of JADES \citep[e.g.,][]{bunker2023} \& FRESCO \citep[e.g,][]{oesch2023} and CEERS \citep[e.g.,][]{finkelstein2023}, respectively, where the values are taken from \cite{hainline2023}. 
The blue triangles denote recent \jwst/NIRSpec observation results for high-redshift lensed galaxies \citep{hsiao2023, williams2023}. 
The grey crosses indicate the photometric candidates with NIRCam publicly available in JADES v1.0 catalog \citep{hainline2023}.  
\label{fig:zspec-Muv}}
\end{figure*}

\setlength{\tabcolsep}{1.5pt}
\begin{deluxetable*}{lcccccccccccc}
\tablecaption{Summary of Spectroscopically-Confirmed Sources at $z\geq8.5$ in UNCOVER}
\tablehead{
\colhead{ID} & \colhead{R.A.} & \colhead{Dec}  & \colhead{$z_{\rm spec}$} & \colhead{F200W}    & \colhead{F444W}       & \colhead{$\mu$} & \colhead{$M_{\rm UV}$} & \colhead{Spec. feature}  & \colhead{$T_{\rm exp.}$} & \colhead{Photo Ref.} & \colhead{Spec Ref.} & \colhead{AGN$?$} \\
      & \colhead{deg}  & \colhead{deg}  &  \colhead{}   &  \colhead{mag}  &  \colhead{mag}  &   &   &  & \colhead{hours}   &  &  &    \\
\colhead{(1)} & \colhead{(2)} & \colhead{(3)} & \colhead{(4)}   & \colhead{(5)}          & \colhead{(6)}               & \colhead{(7)}      & \colhead{(8)}                  & \colhead{(9)}& \colhead{(10)}& \colhead{(11)} & \colhead{(12)}}
\startdata
20466 & 3.640408393 &$ -30.38643761 $& $8.500^{+0.000}_{-0.001}$ & 28.47 & 26.20 &$ 1.3 ^{+0.0}_{-0.0}$ &$ -18.11$& Break, Line & 2.7 (2) & L23 & This &Y$^{\dagger}$\\
10646 & 3.636959984 &$ -30.40636156 $& $8.511^{+0.000}_{-0.001}$ & 25.35 & 24.09 &$ 1.4 ^{+0.0}_{-0.0}$ &$ -21.56$& Break, Line & 2.7 (2) & \nodata & This &  Y/N$\flat$  \\
3686 & 3.617202003 &$ -30.42553381 $& $9.325^{+0.000}_{-0.001}$ & 25.12 & 25.08 &$ 1.6 ^{+0.0}_{-0.0}$ &$ -21.72 $& Break, Line & 2.7 (2) & A23, \tcb{C23} & B23, This & Y/N  \\
22223 & 3.568114512 &$ -30.38305164 $& $9.568^{+0.000}_{-0.001}$ & 28.68 & 28.74 &$ 3.9 ^{+0.3}_{-0.1}$ &$ -17.28 $& Break, Line & 4.4 (4)& \nodata & This &  N \\
31028 & 3.544169292 &$ -30.37031863 $& $9.740^{+0.000}_{-0.001}$ & 27.46 & 27.48 &$ 6.7 ^{+0.1}_{-1.5}$ &$ -17.85 $& Break & 6.9 (3,6) & \nodata & This &  N \\
13151 & 3.592501339 &$ -30.40146429 $& $9.880^{+0.012}_{-0.001}$ & 27.05 & 27.38 &$ 12.8 ^{+0.6}_{-0.8}$ &$ -17.63 $& Break, Line & 11.8 (5,6,7) & Z14 & RB23, This & N \\
26185 & 3.567070796 &$ -30.37786065 $& $10.071^{+0.000}_{-0.001}$ & 27.09 & 27.17 &$ 3.9 ^{+0.5}_{-0.1}$ &$ -18.93 $& Break, Line & 7.1 (1,4) & A23, C23 & G23, This & Y$^{\natural}$ \\
37126 & 3.590110772 &$ -30.35974219 $& $10.255^{+0.001}_{-0.001}$ & 26.85 & 27.41 &$ 1.8 ^{+0.0}_{-0.1}$ &$ -20.01 $& Break & 6.9 (3,4) & A23 & This &  N \\
38766 & 3.513563316 &$ -30.35679963 $& $12.393^{+0.004}_{-0.001}$ & 28.17 & 28.46 &$ 1.5 ^{+0.0}_{-0.0}$ &$ -19.17 $& Break & 4.4 (4) & A23 & W23, This & N \\
13077 & 3.570869325 &$ -30.40158533 $& $13.079^{+0.014}_{-0.001}$ & 27.73 & 28.82 &$ 2.3 ^{+0.0}_{-0.1}$ &$ -19.24 $& Break & 7.4 (5,7) & \nodata & W23, This & N 
\enddata
\tablecomments{
(1): Source ID used in MSA. We also describe the ID used in the photometric catalog \citep{weaver2023} in Appendix \ref{sec:app_ind}. 
(2--3): Source coordinate. 
(4): Spectroscopic redshift ($z_{\rm spec}$) determined by our SED template fitting method (see text).  
(5--6): Total magnitude in NIRCam F200W and F444W filters measured in \cite{weaver2023}. 
(7): Magnification factor based on $z_{\rm spec}$ and the latest lens model, including eight more multiple image systems spectroscopically confirmed in the UNCOVER NIRSpec observations \citep{furtak2023a}. 
(8): Absolute UV magnitude, calculated with the total flux in the NIRCam F150W and F200W filter for the sources at $z_{\rm spec}=8.5$--10 and $z_{\rm spec}>10$, respectively. 
(9): Key spectroscopic features observed in the prism that critically determine the source redshift in our method. ``Line'' denotes the sources with multiple line detections. 
(10): Exposure time in hours. The MSA IDs are also denoted in parentheses. 
(11--12): Reference for photometric and/or spectroscopic results (L23; \citealt{labbe2023}, A23; \citealt{atek2023}, B23; \citealt{boyett2023}, C23; \citealt{castellano2023}, RB23; \citealt{roberts-borsani2023}, G23; \citealt{goulding2023}, W23; \citealt{bwang2023}, Z14; \citealt{zitrin2014}). 
(13): AGN or not. ``Y'' indicates the AGN, while ``N'' indicates no evidence of AGN has been observed in the current data. ``Y/N'' represents the potential AGN sources implied from the emission line properties in prism (Section~\ref{sec:measure}) and UVLF measurements (Section~\ref{sec:uvlf}). \\
$\dagger$ From the broad-line identification in H$\beta$ (see more details in \citealt{kokorev2023b}). ID13556 in \cite{labbe2023}.  \\
$\flat$ See more details in J.~Weaver et al. in (prep.) for line diagnostics and contributions from AGN and star-forming activities. \\ 
$\natural$ Recent deep 1.25 Ms {\it Chandra} observations show the 4.2$\sigma$ detection from ID26185 (\citealt{bogdan2023}; see also \citealt{goulding2023}). 
}
\label{tab:catalog}
\end{deluxetable*}

We perform a template fitting to the NIRSpec prism 1D spectra to measure the source redshift using \texttt{eazy} \citep{brammer2008} implemented in \texttt{msaexp}. 
The \texttt{eazy} code adopts a set of templates added in a non-negative linear combination, allowing us to securely measure the source redshift via any faint emission lines as well as the Lyman-$\alpha$ break feature at $z\gtrsim10$. 
We use the \texttt{corr\_sfhz\_13} subset models\footnote{\url{https://github.com/gbrammer/eazy-photoz/tree/master/templates/sfhz}}
which include redshift-dependent star-formation history (SFH), and dust attenuation. 
We additionally include the best-fit SED template of the \jwst-observed extreme objects of the strong emission line galaxy at $z=8.5$ (ID4590) from \cite{carnall2022} and an obscured AGN at $z=4.50$ in the MACJ0647 lensing cluster \citep{killi2023} to adequately model the potential strong emission lines and obscured AGNs that have been frequently reported in recent NIRSpec studies \citep[e.g.,][]{kocevski2023, harikane2023c, furtak2023c}. 
We include the absorption of the intergalactic medium in the fitting to include the damping Ly$\alpha$ wing effect, especially towards high redshifts \citep[e.g.,][]{curtis-lake2023, umeda2023, heintz2023}. 
We search for the best solution from $\chi^{2}$ minimization over the redshift range of $z=$0.1--20 for all the MOS targets. 
Fixing the best-fit redshift estimate, we also conduct a spline fitting with \texttt{msaexp} to the continuum combined with the single Gaussian for each emission line to evaluate the significance level for each faint emission line at the corresponding wavelength. 

Figure~\ref{fig:spectrum} summarizes the 2D and 1D spectra and the likelihood of the redshift $P(z)$ for ten sources whose redshifts are estimated at $z\geq8.5$ in our analysis. 
All these ten sources show that the likelihood of $z$ below $z=8.0$ $P(z < 8)$ is $\ll$~3e-7. 
This suggests the significance of our spectroscopic redshift confirmation being well beyond the $5\sigma$ level, and we regard these ten sources as the successful spec-$z$ confirmed sources in this paper. 
In Figure~\ref{fig:spectrum}, we also present vertical lines highlighting the wavelengths of the Ly$\alpha$ break and any faint emission lines detected with a signal-to-noise ratio (SNR) $\geq2.5$ via the spline+Gaussian fit to the prism spectrum with the source redshift fixed at the one obtained from the template fitting. 
From 7 out of 10 sources, we detect faint emission lines at secure SNRs ($\geq5$), while the redshift is also constrained for the other three sources via the Ly$\alpha$ break feature. 

Among these ten prism spectra, we detect an unambiguous broad line (BL) H$\beta$ component in ID20466 spectroscopically confirmed at $z=8.50$. The line width of the BL H$\beta$ is estimated to be FWHM $=3439\pm413$ km~s$^{-1}$. 
The Balmer decrement measurement via H$\gamma$/H$\beta$ suggests a heavily dust-obscured nature of $A_{\rm V}= 2.1^{+1.1}_{-1.0}$ with the Small Magellanic Cloud (SMC) dust attenuation law.  
Moreover, it shows a remarkably bright \oiii4363 line, resulting in the dust-corrected \oiii4363/\oiii5008 ratio being 0.32. Such a high ratio cannot be reproduced by typical electron temperatures and electron densities \citep[e.g.,][]{nicholls2020}, while the high ratio is generally observed in local Seyfart galaxies \citep[e.g.,][]{osterbrock1978,dopita1995, nagao2001, baskin2005}. 
From the BL line detection and the extremely high ionization state, we conclude that ID20466 is the AGN, and we refer the reader to the separate paper of \cite{kokorev2023b} for more characterizations and discussions of this source. 

Interestingly, we detect uniquely high ionization emission lines from several other sources. 
For example, ID10646 shows a remarkable number of emission lines with secure SNRs, including \civ1549 and \heii1640, similar to the AGN candidate of GNz11 \citep{maiolino2023c}. 
\tcb{A detailed UV--optical line study (H.~Treiber et al. in prep.) suggests that ID10646 falls on the AGN regime in the equivalent width (EW) -- line relation among the \ciii1907,1909, \civ1549, and \heii1640 lines, based on the rest-frame UV-optical line diagnostic for star-forming activity and AGNs \citep[e.g.,][]{feltre2016,hirschmann2019}. 
This indicates that ID10646 is a plausible AGN candidate.} 
\tcb{ID3686 also shows a unique excitation feature.}
The ratio of \oiii5008/H$\beta$ is observed to be $16.3^{+21.8}_{-5.9}$, which exceeds the maximum value of $\sim10$ observed in recent NIRSpec studies for galaxies at $z\sim2$--9 and falls in the AGN regime in the \nii, [S\,{\sc ii}], and [O\,{\sc i}] BPT diagrams \citep[e.g.,][]{sanders2023}. 
\tcb{ID3686 falls in the AGN regime also in the Mass-Excitation diagnostic \citep[e.g.,][]{juneau2014} with the stellar mass estimate of $\sim10^{9}M_{\odot}$ for this object \citep{boyett2023}.} 
Given the potential self-subtraction due to its extended morphology and the three-shutter nod method (see Figure~\ref{fig:cutout}), we also perform a global background subtraction by using a nearby empty shutter and confirm a similarly high ratio of $11.2^{+4.2}_{-2.4}$ in the central shutter. 
While the high \oiii5008/H$\beta$ ratio is also induced by the shock excitation \citep[e.g.,][]{kewley2013, hirschmann2022}, we also confirm a high ratio reaching out to $17.5^{+17.5}_{-5.8}$ in the spectrum of the Southern East shutter with the global subtraction method, where the emission is more dominated by the compact component in the NIRCam map. 
These results may also suggest that the uniquely high \oiii5008/H$\beta$ ratio in ID3686 is the strong radiation from an AGN. 
Note that both ID10646 and ID3686 are spatially resolved in the NIRCam filters. However, the BL AGNs have also been identified in spatially-resolved sources at $z\simeq4-7$ \citep[e.g.,][]{harikane2023c}, and the presence of the AGN does not always require a point source morphology, depending on the contrast between the host galaxy and AGN. 
\tcb{Although a top-heavy IMF in high-$z$ galaxies is also an interesting and plausible possibility to explain the unique excitation features, the top-heavy IMF itself also still needs confirmation from observations.
In short, although the origins of unique excitation features observed in ID10646 and ID3686 are not conclusive for now, AGNs naturally fit with those unique excitation features, and we regard those two sources as potential AGN candidates in the following sections.}

In Table~\ref{tab:catalog}, we summarize our spectroscopic redshift estimates $z_{\rm spec}$ for our ten spec-$z$ confirmed sources, together with basic source properties and the implications of the AGN. We further discuss the potential AGN interpretation of ID10646 and ID3686 in Section~\ref{sec:uvlf}. 
In Figure~\ref{fig:zspec-Muv}, we summarize $M_{\rm UV}$ as a function of redshift for our spec-
$z$ confirmed sources, together with the photometric and spectroscopic sample in the literature at $z\gtrsim8$. Owing to the deep survey layer efficiently explored by the gravitationally lensing effect, our UNCOVER sources increase the sample in a high-redshift ($z\gtrsim9$) and faint ($M_{\rm UV}\lesssim-19$) regime by a factor of 3.

\begin{figure*}
\begin{center}
\includegraphics[trim=0cm 0cm 0cm 0cm, clip, angle=0,width=1.\textwidth]{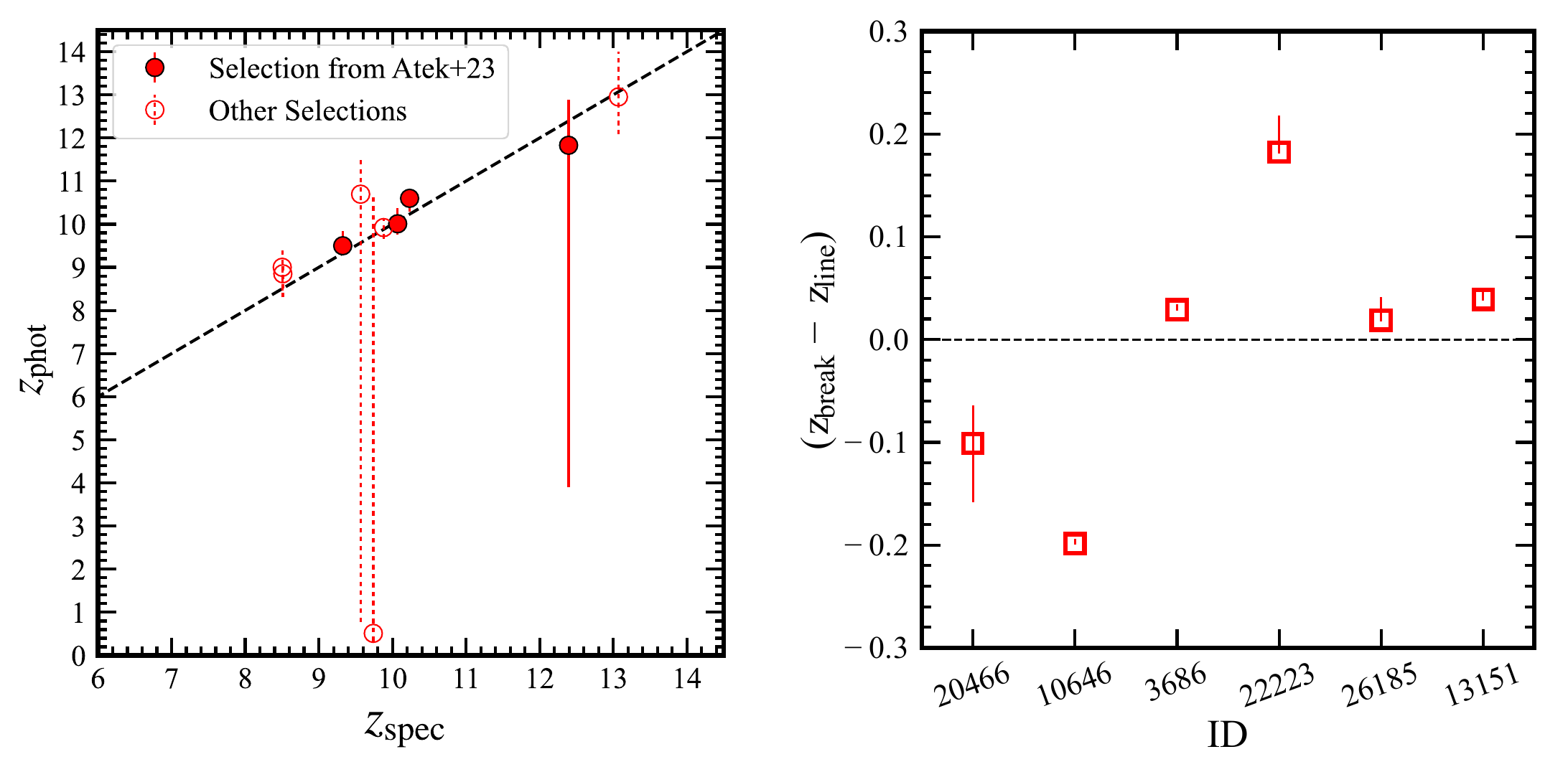}
\end{center}
\vspace{-0.4cm}
 \caption{
\textbf{\textit{Left:}} 
Comparison between $z_{\rm phot}$ and $z_{\rm spec}$ for our spec-$z$ confirmed sources.  The red-filled and open circles represent the sources selected from \cite{atek2023} (hereafter A23) and other selections, respectively (see Section~\ref{sec:measure}). 
The error bars show the $1\sigma$ error range for the A23 sample, while the other sources show the 2$\sigma$ error range given non-homogeneous selection criteria adopted in the MSA target selection process (Section~\ref{sec:nircam}).   
\textbf{\textit{Right:}}
Comparison between the redshift estimates based on lines ($z_{\rm line}$) and the Ly$\alpha$ break ($z_{\rm break}$) for six sources whose multiple emission lines are successfully detected in the prism spectrum. 
Two sources (ID20466, ID10646) show $z_{\rm break}-z_{\rm line}< -0.1$, indicating the presence of the non-zero fluxes at the blueward of the Ly$\alpha$ emission in the prism spectra and the ionized bubbles around these two sources (Section~\ref{sec:bubble}). 
\label{fig:zph-zspec}}
\end{figure*}

\subsection{$z_{\rm phot}$ vs $z_{\rm spec}$}
\label{vs_zph}

In the left panel of Figure~\ref{fig:zph-zspec}, we compare the $z_{\rm phot}$ and $z_{\rm spec}$ estimates for the spec-$z$ confirmed sources. 
For the four sources presented in A23 (red filled circles), we adopt the $z_{\rm phot}$ estimates from A23 with the 1$\sigma$ error range. 
For the other six sources (red open circles), we use the $z_{\rm phot}$ estimates from the \texttt{eazy} fitting with the default \texttt{corr\_sfhz\_13} template set \citep{bwang2024a}. For the latter six sources, we show the 2$\sigma$ error range, given the non-homogeneous selection criteria adopted in the MSA target selection process (Section~\ref{sec:nircam}). 
We find that all $z_{\rm phot}$ estimates agree with $z_{\rm spec}$ within the $\sim1$-2$\sigma$ error ranges. 
Among the sources presented in A23, there are no other sources included in our MSA design apart from the four sources, resulting in the success ratio of the spec-$z$ confirmation being 100\% ($=4/4$) for the A23 sample. 
This high confirmation rate is consistent with previous systematic NIRSpec follow-up studies for NIRCam-selected high-redshift candidates at $z\gtrsim9$ in the CEERS survey ($\simeq90$\%; \citealt{fujimoto2023a}; \citealt{arrabal-haro2023b}), validating the classical high-redshift galaxy selections based on the $z_{\rm photo}$ estimates and/or the dropout technique. 
Interestingly, the A23 sample shows the offset of the redshift $\Delta z (\equiv z_{\rm phot}-z_{\rm spec}) \in[-0.4:+0.4]$, which is in contrast to the previous NIRSpec follow-up studies showing a trend of the overestimate in $z_{\rm phot}$ typically $\Delta z\simeq+0.5$ and maximally $\sim+1$--2, regardless of the choice of the $z_{\rm phot}$ estimates from different literature \citep{fujimoto2023a, arrabal-haro2023b, hainline2023}. 
The overestimate of $z_{\rm phot}$ is likely because of the softened Ly$\alpha$-break shape routinely identified in the NIRSpec spectra for high-redshift galaxies with possible causes of the IGM Ly$\alpha$ absorption, the intrinsic SED shape, and/or the additional Damping Ly$\alpha$ Absorbing systems (DLAs) \citep[e.g.,][]{curtis-lake2023, arrabal-haro2023b, hsiao2023, umeda2023, heintz2023}. 
In addition to the deep NIRCam blue (F115W, F150W) filters taken in UNCOVER ($5\sigma\simeq30$~mag) that are deeper than those taken in CEERS by $\simeq 1$~mag \citep{finkelstein2023}, one of the most strict selection criteria is adopted in the A23 selection, including the sharp Ly$\alpha$-break color of $>1.5$~mag and the consistent redshift solutions from different SED fitting codes. 
The presence of the deep blue filters and the strict sample selection may be plausible reasons that mitigate the overestimation of $z_{\rm phot}$ in the A23 sample.

\subsection{$z_{\rm line}$ vs $z_{break}$}
\label{vs_zbreak}

\begin{figure*}
\begin{center}
\includegraphics[trim=0cm 0cm 0cm 0cm, clip, angle=0,width=1.\textwidth]{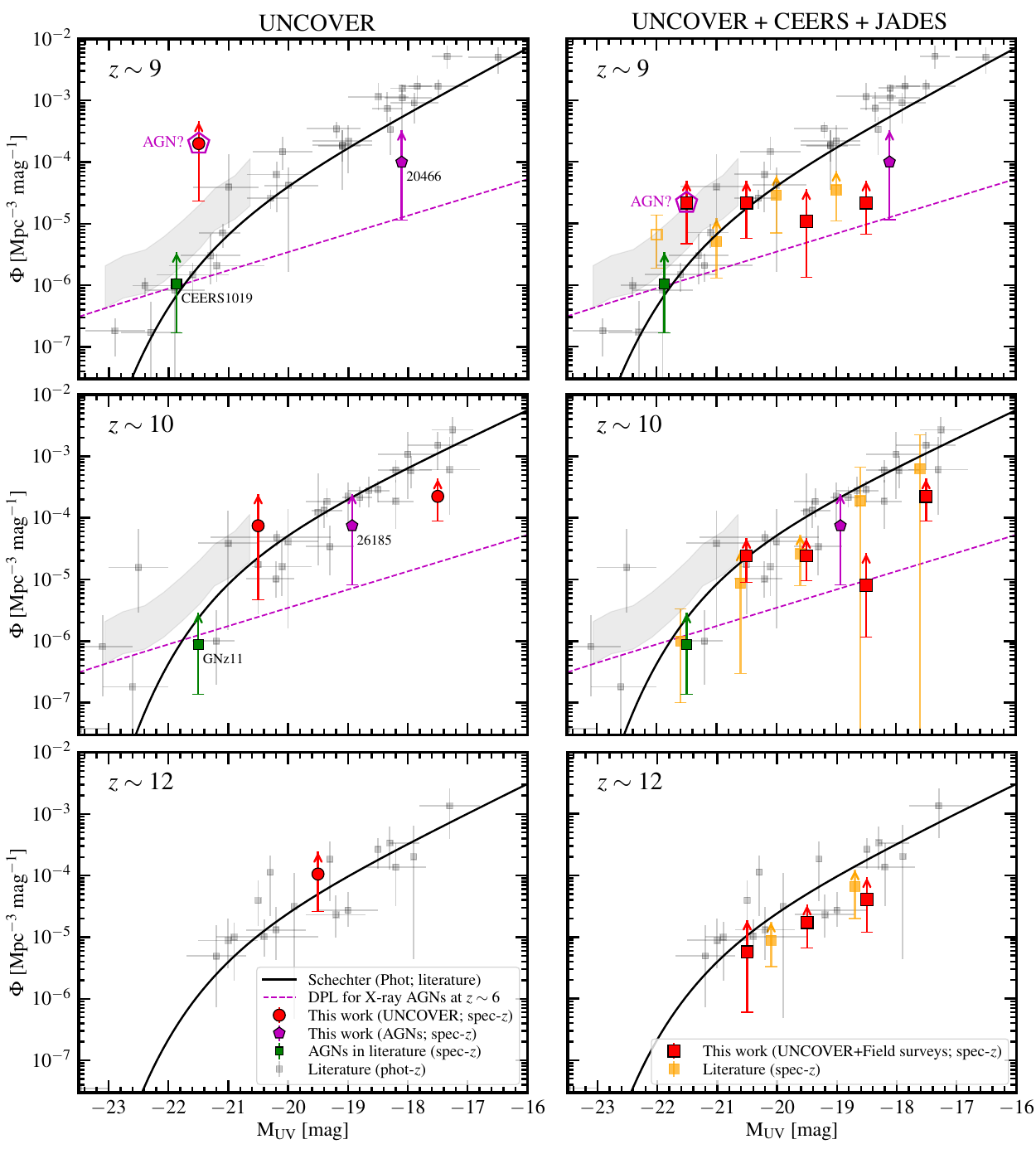}
\end{center}
\vspace{-0.6cm}
 \caption{
Constraints on the UV luminosity function at $z\sim9$, $z\sim10$, and $z\sim12$. 
The magenta pentagons represent the AGN sources from the BL (ID20466) and X-ray (ID26185) detection, and the red circles present the other eight sources in our spec-$z$ confirmed sample in UNCOVER. 
The red squares denote the spec-$z$ confirmed non-obvious AGN sources from UNCOVER, CEERS, and JADES GOODS-S, where we do not include several sources in the $z=8.7$ overdensity reported in the CEERS field (see text).  
The grey shaded region and grey squares show the previous photometric measurements, and the black solid curve denotes the best-fit Schechter function estimated in \cite{perez-gonzalez2023}. 
The orange squares show the recent spectroscopic measurements \citep{harikane2023c}. 
The green squares denote the recent \jwst-observed bright objects of GNz11 \citep{maiolino2023b} and CEERS1019 \citep{larson2023} reported as AGNs. 
The magenta dashed curve presents the best-fit Double power law (DPL) function for $z\sim6$ AGNs \cite{giallongo2019}.  
The open magenta pentagon remarks the two possible AGN sources at different redshifts that show uniquely high ionization emission lines such as \niv1749, \civ1549, and \heii1640 (ID10646 at $z=8.51$) and high \oiii5008/H$\beta$ ratio of $>10$ (ID3686 at $z=9.33$), where the excess is not caused by an overdensity but by their uniquely UV-bright properties. 
\label{fig:uvlf}}
\end{figure*}

Among our spec-$z$ confirmed sources, some are detected with multiple emission lines, making their $z_{\rm spec}$ estimates very secure. 
On the other hand, others without the multiple emission line detection mainly rely on the Ly$\alpha$ break feature, which may still have uncertainty in their $z_{\rm spec}$ estimates. 
Motivated by this, we compare the line-based redshift estimate ($z_{\rm line}$) and the Ly$\alpha$-break-based redshift estimate ($z_{\rm break}$).
We use ID20466, ID10646, ID3686, ID22223, ID26185, and ID13151, which show multiple emission line detection (Figure~\ref{fig:spectrum}) and are suitable for this experiment. 
For $z_{\rm line}$, we mask the $\pm0.1\mu$m range from the observed Ly$\alpha$ wavelength in the 1D spectra. For $z_{\rm break}$, we mask all wavelengths with $\lambda \geq 3.0\mu$m and the $\pm0.05\mu$m ranges from the observed wavelengths of detected emission lines in the 1D spectra. 
We then rerun the same template fitting to the masked spectra as Section~\ref{sec:measure} and derive $z_{\rm line}$ and $z_{\rm break}$. 

In the right panel of Figure~\ref{fig:zph-zspec}, we compare our $z_{\rm line}$ and $z_{\rm break}$ estimates.  We find that the offset between the line and Ly$\alpha$ break-based redshifts $\Delta z' (\equiv z_{\rm line}-z_{\rm break}) \in [-0.2:+0.2]$. This indicates that the redshift estimate based on the Ly$\alpha$ break feature alone may still have the uncertainty of $\pm0.2$. Given that we perform this experiment only with the sources with multiple significant emission line detections, which are mostly equal to the best SNR spectra, the offset could be even worse than $\pm0.2$ with lower SNR data. 
This is important for the design of future follow-up spectroscopy based on the redshift estimates with NIRSpec/prism, especially when using instruments whose redshift coverage of targeting emission lines can be narrower than this potential redshift uncertainty, such as ALMA. 

Individually, ID22223 shows $z_{\rm break}$ overestimated by $\sim$0.2. 
The 1D spectrum clearly shows the softened shape of the Ly$\alpha$ break, 
which is the natural cause of the overestimate of $z_{\rm break}$. 
ID3686, ID26185, and ID13151 also show slight overestimates in $z_{\rm break}$ ($\Delta z' < 0.05$). In the 1D spectra of these three sources, we identify the softened shape in the Ly$\alpha$ break more or less similar to that of ID22223. 
These results indicate that the slight overestimate of $z_{\rm break}$ in these three sources is also caused by the similar softened shape of the Ly$\alpha$ break, while the clear cut-off of the Ly$\alpha$ break mitigates the effect. 
Therefore, an overestimation of $z_{\rm break}$ does not always \tcb{severely occur due to the softened Ly$\alpha$ break,}
depending on the SNR of the spectrum. 

In ID20466 and ID10646, on the other hand, the $z_{\rm break}$ values are underestimated by $\sim$0.1--0.2. 
Their 1D spectra show that the Ly$\alpha$ line and continuum continue down to shorter wavelengths than the rest-frame 1216${\rm \AA}$, indicative of an enhanced transmission of the Ly$\alpha$ line and continuum due to the presence of the so-called proximity zone, which has been often observed around high-redshift luminous quasars \citep[e.g.,][]{eilers2017}.  
We further discuss the proximity zone around ID20466 and ID10646 in Section~\ref{sec:bubble}.

\section{UVLF at $\lowercase{z}\gtrsim9$ and implications of AGN contributions}
\label{sec:uvlf}

\subsection{UVLF from UNCOVER}
\label{sec:uvlf_uncover}

We calculate the UV LFs at $z\gtrsim9$ using our spec-$z$ confirmed sources. 
Note that we could not assign all the $z\gtrsim9$ photometric candidates slits in the MSA design, and thus our measurements provide only the lower limits. In the same manner as \cite{harikane2023b}, we divide our sample into three redshift bins at $z_{\rm spec}=8.5-9.5$, $9.5-11.0$, and $11.0-13.5$. To simplify the calculation, we adopt a top-hat function for the volume calculation according to the redshift bin. 
For the survey area, since the MSA footprints are \tcb{restricted} to the high-priority targets originally identified from the NIRCam observation around the primary and two sub-cluster regions in A2744, we use the NIRCam mosaic for the primary UNCOVER area ($\sim28$~arcmin$^{2}$; \citealt{bezanson2022}) and obtain the effective survey area by applying the magnification correction in the same manner as \cite{atek2023}. 
\tcb{
Note that two objects (ID3686, ID10646) slightly outside the primary UNCOVER area still fall within our MSA footprint near the edge. However, including all outside NIRCam areas taken in other programs (e.g., GLASS) would be inappropriate as most are not covered by our MSA footprint due to the restriction. Additionally, our MSA footprint does not cover about 40\% of the primary UNCOVER area itself. Given these factors, using the 28~arcmin$^{2}$ area provides a reasonable balance for our survey volume estimation.
}
The uncertainty is calculated from Poisson errors with the values presented in \cite{gehrels1986}, where we take the cosmic variance into account, following \cite{trapp2020}. 
No completeness correction is applied, which makes our lower limit estimates conservative. 
Our full UVLF measurements using the photometric sample leveraged by the success ratio of the spec-$z$ confirmation are presented in \cite{chemerynska2024}.

The left panel of Figure~\ref{fig:uvlf} shows our UVLF measurements for galaxies (red circles; $N=8$) and AGNs (magenta pentagons; $N=2$). For comparison, we also present the previous photometric UVLF measurements (grey squares) taken from the literature \citep{harikane2023, donnan2023, bouwens2022d, leethochawalit2023, bouwens2021, stefanon2019, bowler2020, mcleod2016, oesch2018, bagley2022a, morishita2018, morishita2022, castellano2022, naidu2022c, finkelstein2022aa, finkelstein2022b, leung2023, perez-gonzalez2023, franco2023} and the best-fit Schechter function (black curve) presented in \cite{perez-gonzalez2023}. 
We also show two bright \jwst-observed sources, GNz11 and CEERS1019, that are argued to be AGNs in the literature \citep[e.g.,][]{larson2023, maiolino2023a}. For GNz11, we calculate the survey area with the entire CANDELS field (736~arcmin$^{2}$; \citealt{bouwens2021}). For CEERS1019, we calculate the survey area from the CANDLES fields of GOODS-N, EGS, UDS, and COSMOS, where the original spectroscopic sample was selected \citep{larson2022}. For volume calculations, we assume $\Delta z=1.0$ in the same manner as \cite{oesch2018} for both sources.
The uncertainty is estimated by the Poisson error and the cosmic variance. 

We find that our measurements for both galaxies and AGNs are consistent with the previous photometric measurements, except for the brightest $M_{\rm UV}$ bin in the $z\sim9$ UVLF. 
This exceptional data point consists of our two brightest sources, ID3686 at $z=9.33$ and ID10646 at $z=8.51$. These two sources are located at different redshifts, indicating that the excess from the previous UVLF measurements is not caused by an overdensity but by their uniquely UV bright properties. Interestingly, both of these sources show some implications as AGNs from the identification of the high ionization emission lines such as \niv1749, \civ1549, and \heii1640 and the uniquely high \oiii5008/H$\beta$ ratio of $>10$ (Section~\ref{sec:measure}). 
Although both of these sources are spatially resolved in the NIRCam filters, 
many BL AGNs have been identified in spatially-resolved sources in recent NIRSpec studies \citep[e.g.,][]{harikane2023c, larson2023}. Even if the AGN contribution to the total UV luminosity is 50\%, the $M_{\rm UV}$ value moves towards the bright end by $\sim$0.7~mag, which may easily impact the bright-end shape of the UVLF. 
These UVLF measurements offer another independent implication of the potential AGN nature of these UV brightest sources. 

In the left panel of Figure~\ref{fig:uvlf}, we also show the AGN LF estimated at $z\sim6$ for X-ray AGNs (the magenta dashed line; \citealt{giallongo2019}). This X-ray AGN LF shows an excellent agreement with the lower boundaries obtained from the BL AGN of ID20466 at $z=8.50$ and the X-ray luminous AGN (ID26185) at $z=10.07$. This indicates that a comprehensive AGN LF at $z\gtrsim9$ could have a comparable, even higher amplitude than the previous measurement at $z\sim6$. 
We summarize our lower limit constraints on UVLFs of galaxies and AGNs in Table~\ref{tab:uvlf} and Table~\ref{tab:uvlf2}, respectively.

\setlength{\tabcolsep}{10pt}
\begin{deluxetable}{lcc}
\tablecaption{Spectroscopic Constraints on $z\simeq9$--12 UVLFs}
\tablehead{\colhead{$M_{\rm UV}$}  & \colhead{$\Phi$ (UNCOVER)} & \colhead{$\Phi$ (UNCOVER+Fields)} \\
   \colhead{[AB mag]}          & \multicolumn{2}{c}{[$10^{-5}$ Mpc$^{-3}$~dex$^{-1}$]} \\
        (1)   &    (2)          &          (3)     }
\startdata
\multicolumn{3}{c}{$z\sim9$} \\ \hline 
$-21.5$ & $>19.9_{-17.6}^{+25.9}$ &$>2.16_{-1.69}^{+2.81}$ \\
$-20.5$ & \nodata & $>2.16_{-1.59}^{+2.81}$ \\
$-19.5$ & \nodata & $>1.08_{-0.95}^{+2.49}$ \\
$-18.5$ & \nodata & $>2.16_{-1.49}^{+2.81}$ \\
\hline \multicolumn{3}{c}{$z\sim10$} \\  \hline 
$-20.5$ & $>7.45_{-6.98}^{+17.1}$ &$>2.43_{-1.53}^{+2.34}$\\
$-19.5$ & \nodata & $>2.43_{-1.47}^{+2.34}$ \\
$-18.5$ & \nodata & $>0.81_{-0.69}^{+1.86}$ \\
$-17.5$ & $>22.4_{-13.5}^{+21.6}$ &$>22.4_{-13.5}^{+21.6}$\\
\hline \multicolumn{3}{c}{$z\sim12$} \\  \hline 
$-20.5$ & \nodata & $>0.69_{-0.62}^{+1.58}$ \\
$-19.5$ & $>10.6_{-7.88}^{+13.8}$ &$>2.06_{-1.27}^{+1.99}$ \\
$-18.5$ & \nodata & $>4.08_{-2.90}^{+5.31}$ \\
\enddata
\tablecomments{
(1): Absolute UV magnitude.  
(2): UVLF constraints with spec-$z$ confirmed sources in UNCOVER,  except for two AGNs (see text), resulting in $N=8$.  
(3): UVLF constraints with spec-$z$ confirmed sources in UNCOVER, CEERS, and JADES, except for two AGNs \citep{larson2023, maiolino2023b} and the sources in the $z=8.7$ overdensity in the CEERS field reported in the literature \citep{larson2022}, resulting in $N=23$. 
Errors and upper limits are 1$\sigma$, evaluated with the Poisson uncertainty \citep{gehrels1986} and the cosmic variance \citep{trapp2020}. 
}
\label{tab:uvlf}
\end{deluxetable}

\setlength{\tabcolsep}{15pt}
\begin{deluxetable}{lcc}
\tablecaption{Spectroscopic Constraints on $z\simeq$9--10 AGN LFs}
\tablehead{\colhead{Redshift}  & \colhead{$M_{\rm UV}$}  & \colhead{$\Phi$ (AGN)} \\
                          &       [AB mag]      &   [$10^{-5}$ Mpc$^{-3}$~dex$^{-1}$]}
\startdata
$z\sim9$  & $-18.1$ & $>9.96_{-8.81}^{+22.9}$ \\
$z\sim10$ & $-18.9$ & $>7.45_{-6.63}^{+17.1}$  \\
\enddata
\tablecomments{
Same as Table~\ref{tab:uvlf}, but for AGN LFs constrained from two spec-$z$ confirmed AGNs of ID20466 and ID26185. 
}
\label{tab:uvlf2}
\end{deluxetable}

\begin{figure*}
\begin{center}
\includegraphics[trim=0cm 0cm 0cm 0cm, clip, angle=0,width=1.\textwidth]{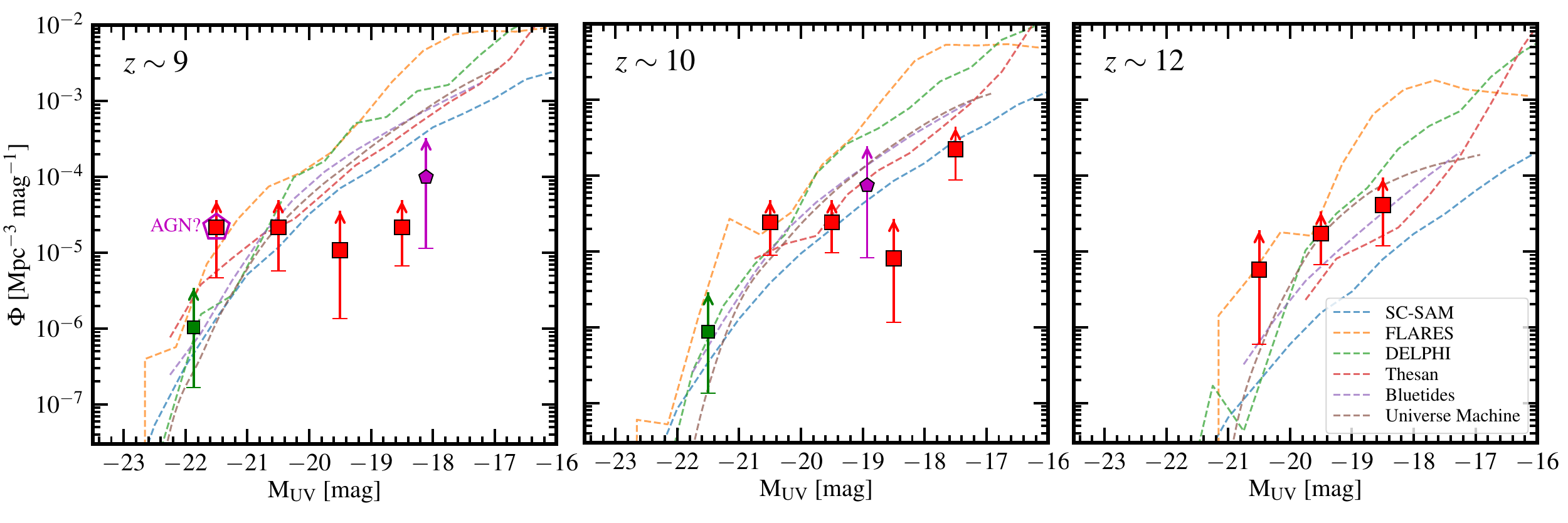}
\end{center}
\vspace{-0.6cm}
 \caption{
Comparison of the luminosity functions with the theoretical predictions in the literature at $z\sim9$, $z\sim10$, and $z\sim12$.  
The symbols are the same as Figure~\ref{fig:uvlf}, 
The color dashed lines denote the theoretical predictions of SC-SAM \citep{yung2019a}, FLARES \citep{lovell2022, vijayan2021, wilkins2023}, DELPHI \citep{dayal2014, mauerhofer2023}, Thesan \citep{kannan2022}, Bluetides \citep{wilkins2017}, and Universe Machine \citep{behroozi2020}.  
In each redshift range, our lower limit estimate in the brightest $M_{\rm UV}$ challenges several theoretical predictions, confirming the earlier arguments of the high abundance of UV-bright ($M_{\rm UV} \lesssim -20$) galaxies in previous photometric studies in a spectroscopic manner at $z\sim9$--12. 
\label{fig:uvlf_model}}
\end{figure*}

\subsection{UVLF from UNCOVER+CEERS+JADES}
\label{sec:uvlf_comb}

To benefit from the complementary survey layers enabled by the lensing cluster surveys and the general field surveys, we also evaluate the UVLF together with public spec-$z$ confirmed sources at $z\geq8.5$ in recent \jwst/NIRSpec MSA studies \citep[e.g.,][]{tang2023, fujimoto2023a, bunker2023, bunker2023b, hainline2023, arrabal-halo2023a, arrabal-haro2023b, harikane2023c, hsiao2023}. 
We include the sources spectroscopically confirmed at $z\geq8.5$ in two general field surveys of CEERS \citep[e.g.,][]{finkelstein2023, arrabal-haro2023b, fujimoto2023} and JADES in GOODS-S (hereafter JADES-GS) \citep[e.g.,][]{bunker2023b, curtis-lake2023, robertson2023, hainline2023} to our spectroscopically confirmed sources in UNCOVER. 
We also use the sources spectroscopically confirmed in a \jwst\ DDT follow-up observation with NIRSpec/prism MSA, which primarily aims to confirm a remarkably UV-bright galaxy candidate at $z\sim16$ \citep{arrabal-halo2023a}. We obtain 12 and 7 spec-$z$ confirmed sources from CEERS (+DDT) and JADES-GS, respectively, covering a wide $M_{\rm UV}$ range of $\in[-22.2:-18.0]$. In conjunction with our UNCOVER sample, our final spec-$z$ sample in this analysis results in a total number of 29. 

We calculate the survey volume as follows. 
In JADES-GS, the NIRSpec observations had small dithers ($<1''$) with one pointing, and the NIRSpec pointing center was not re-optimized to the late addition of the high-priority NIRCam sources \citep{bunker2023b}.  
Therefore, we regard the JADES-GS NIRSpec observations as a pure general field survey almost with a single NIRSpec pointing and adopt the NIRSpec Field-of-View (FoV) of 9~arcmin$^{2}$.\footnote{\url{https://jwst-docs.stsci.edu/jwst-near-infrared-spectrograph\#gsc.tab=0}} 
In CEERS, the NIRSpec observations were initially designed with six pointings, where one pointing was added from the DDT NIRSpec observations (\#2750: PI, P.~Arrabal-Haro; \citealt{arrabal-haro2023b}) for the $z\sim16$ candidate \citep{donnan2023}. 
Among the original six pointings in CEERS, the prism observations in two pointings were affected by the electrical short and thus rescheduled \citep{arrabal-haro2023b}. 
Although the pointing centers of the rescheduled two NIRSpec observations were optimized to maximize the yield of the NIRCam-selected high-redshift galaxy candidates,
we assume that the potential bias from this optimization for the targets is insignificant, given the ratio to the total number of the pointings (= 2/9 $\lesssim20\%$). 
Since the primary target in the DDT observation is the $z\sim16$ candidate, we assume that the potential bias in the DDT observation for other $z\gtrsim9$ galaxies within the NIRSpec FoV is also minimal. 
In a similar manner as \cite{harikane2023b}, we count the area overlapped between the FoVs of the NIRSpec and NIRCam in CEERS and obtain 37.5~arcmin$^{2}$. 
We add the survey areas from JADES-GS and CEERS to that of UNCOVER, and re-derive the UVLFs with the three redshift bins at $z_{\rm spec} =$ 8.5--9.5, 9.5--11.0, 11.0--13.5. 
We do not include CEERS-D28 at $z=8.763$, CEERS1025 at $z=8.715$, CEERS1019 at $z=8.679$, CEERS80083 at $z=8.638$, and CEERS1029 at $z=8.610$ \citep{arrabal-haro2023b} in our estimates, as the $z=8.7$ overdensity has been reported in literature \citep[e.g.,][]{larson2022, castellano2023, harikane2023c}.  

In the right panel of Figure~\ref{fig:uvlf}, we present the UVLF measurements with the spec-$z$ confirmed sources in UCNOVER, CEERS, and JADES. 
For comparison, we also present the recent NIRSpec spec-$z$ based measurements (the orange squares; \citealt{harikane2023c}). 
We confirm that our measurements are generally consistent with the previous spec-$z$ based measurements and improve the lower constraints owing to the comprehensive spec-$z$ sample, including the lensing cluster and general field surveys. 
We find that the brightest $M_{\rm UV}$ bin in the $z\sim9$ UVLF still shows the excess from the previous photometric UVLF measurements beyond the errors, which still consists of ID10646 and ID3686. This indicates that the excess is unlikely explained by the cosmic variance. 
A similar excess has also been reported in recent \jwst\ studies \citep[e.g.,][]{castellano2023, harikane2023c}, and this has been interpreted as galaxy overdensity. 
However, we do not include the sources in the $z=8.7$ overdensity \citep{larson2022} in our measurements. Besides, ID3686 and ID10646 are at different redshifts. 
Thus, the excess is more likely caused by their uniquely UV-bright properties than the excess in abundance, offering independent implications that they are the AGNs, as discussed in Section~\ref{sec:uvlf_uncover}. 
In Table~\ref{tab:uvlf2}, we also summarize the UVLF measurements using the spec-$z$ sources from UNCOVER, CEERS, and JADES.

\subsection{Comparison with Models}
\label{sec:uvlf_model}

We compare our UVLF measurements with theoretical predictions. 
In Figure~\ref{fig:uvlf_model}, we show theoretical predictions of the UVLF at $z\sim9$, $z\sim10$, and $z\sim12$, together with our UVLF measurements using the spec-$z$ confirmed sources in UNCOVER, CEERS, and JADES. 
We find that our lower limit estimate in the brightest $M_{\rm UV}$ bin challenges some theoretical predictions in every redshift range. 
This indicates that we confirm the earlier reports of the high abundance of UV-bright ($M_{\rm UV}\lesssim-20$) galaxies argued in the photometric studies \citep[e.g.,][]{adams2022, atek2022, atek2023, bouwens2022c, bradley2022, castellano2022, donnan2023, finkelstein2022b, finkelstein2023, harikane2023, labbe2022, morishita2022, naidu2022, yan2022, williams2023, austin2023, leung2023} in a spectroscopic manner at $z\sim9$--12. 
Several possible scenarios for the high abundance have been discussed, including preferential detection of galaxies upscattered compared to the main sequence, lower dust attenuations with increasing redshift, a top-heavy initial mass function or even UV luminosity contribution from an AGN component \citep[e.g.,][]{ferrara2022,inayoshi2022, pacucci2022, naidu2022c, finkelstein2023, bouwens2022d, harikane2023c}.
Given that the most stringent lower limit obtained at $z\sim9$ is dominated by possible AGN sources (Section~\ref{sec:uvlf_uncover} \& \ref{sec:uvlf_comb}), it seems increasingly plausible that the bright end of the UV LF is shaped by sources where the UV luminosity is contributed by both star formation and black hole accretion. 
Although further observational evidence is necessary to conclude whether those possible sources are truly AGNs or not, such observations will prove crucial in baselining theoretical models and shedding light on black hole seeding \& growth at these early epochs.

\subsection{AGN fraction}
\label{sec:agn_frac}

\begin{figure}
\begin{center}
\includegraphics[trim=0cm 0cm 0cm 0cm, clip, angle=0,width=0.5\textwidth]{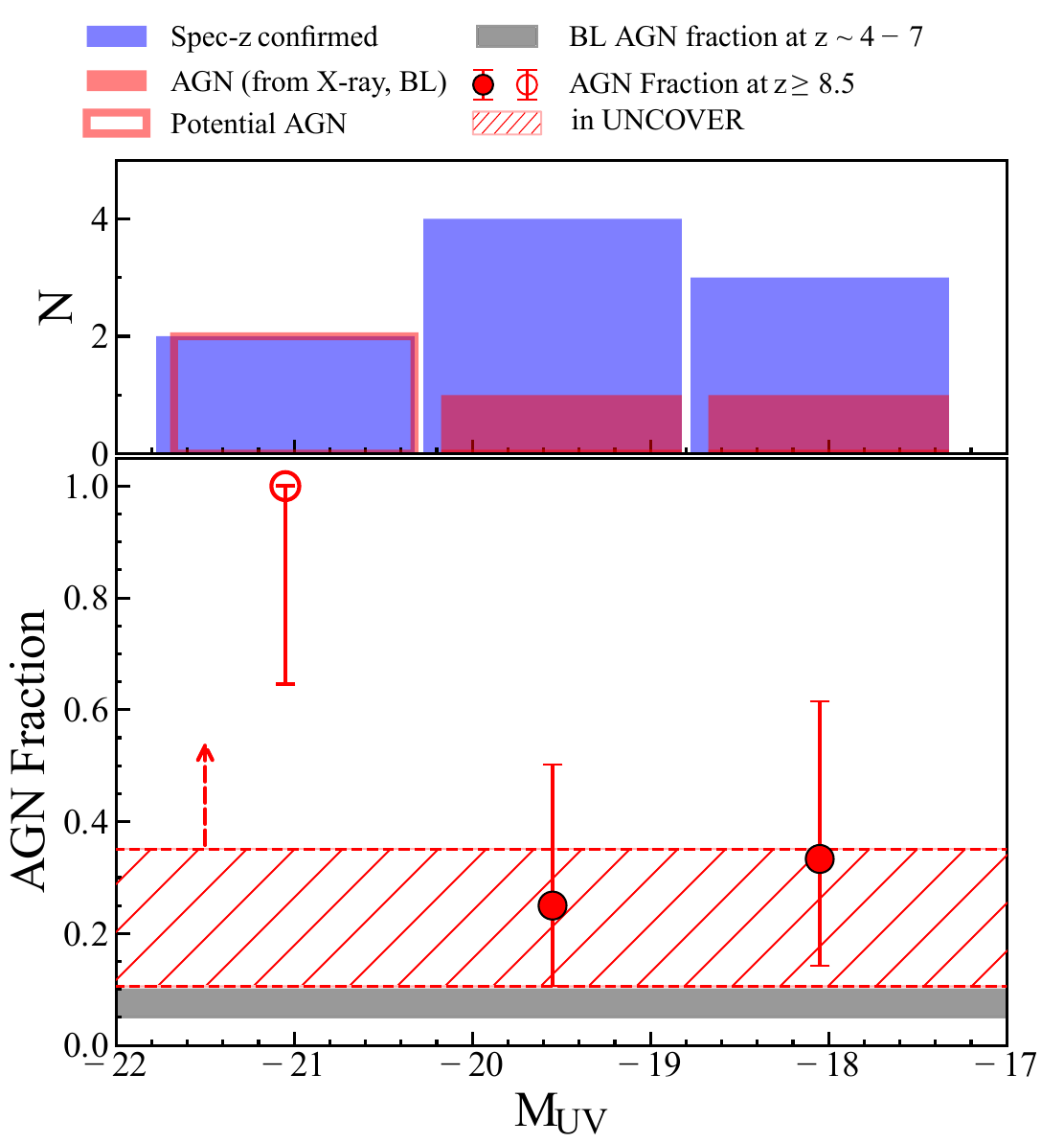}
\end{center}
\vspace{-0.2cm}
 \caption{
AGN fraction at $z\geq8.5$. 
The blue and red histograms present the number of AGNs and the sources that are spectroscopically confirmed in our studies at each $M_{\rm UV}$ bin.
The histogram with a red outline represents the two possible AGN sources of ID10646 and ID3686 that fall in the brightest $M_{\rm UV}$ bin. 
The red circles indicate the AGN fraction for the spec-$z$ confirmed UNCOVER sources at $z\geq8.5$, where we show the brightest $M_{\rm UV}$ bin in the case that both \tcb{AGN candidates are truly AGNs}.  
The error bars represent the confidence intervals for the binomial proportion, derived using the Jeffreys interval at 1$\sigma$. 
The red-hatched area shows the 1$\sigma$ range for the AGN fraction over the entire $M_{\rm UV}$ range ($=2/10$). This may be a lower limit, as more sources, including ID10646 and ID3686, might be confirmed to be AGNs in future observations. 
The black shaded area denotes the recent reports of $\simeq 5-10\%$ from the BL AGN identifications in recent NIRSpec studies at $z\simeq4$--7 \citep{harikane2023b, maiolino2023c}.  
\label{fig:agn_frac}}
\end{figure}

We investigate the AGN fraction at $z\geq8.5$. 
Note that here we count the spec-$z$ confirmed sources in UNCOVER alone, 
since the homogeneous data and sensitivity are required to examine the AGN fraction. 
Figure~\ref{fig:agn_frac} presents the AGN fraction as a function of $M_{\rm UV}$ from our spec-$z$ sample. 
We calculate the 1$\sigma$ uncertainty from the confidence intervals for the binomial proportion, derived using the Jeffreys interval. 
The red-hatched area indicates the 1$\sigma$ range of the AGN fraction over the entire $M_{\rm UV}$ range with our secure AGNs from the BL and X-ray detections ($=2/10$), while we also show the possible AGN fraction in the brightest $M_{\rm UV}$ bin which consists of the possible AGN sources of ID10646 and ID3686.  
For comparison, we also show the BL AGN fraction of $\approx$5--10\% estimated at $z\simeq4$--7 in the recent \jwst/NIRSpec studies \citep{harikane2023b, maiolino2023c}. 

We find that our results suggest a relatively high AGN fraction of $> 10$--35\% compared to the previous measurements from the BL AGNs. 
\tcb{Although a potential bias in the MSA target selection might exist (e.g., robust photo-$z$ sources),}
our higher AGN fraction than the previous measurements can be interpreted as a more comprehensive approach adopted in the AGN identification than the BL identification alone. 
Since the weak-line quasars have also been identified at $z>6$ \citep[e.g.,][]{andika2020, fujimoto2022}, the BL approach can be hampered by the observational bias for the sources whose equivalent width of BL is high ($\approx$ high $M_{\rm BH}/M_{\rm star}$) and/or whose BL width is sufficiently broad ($\approx$ more massive $M_{\rm BH}$) to be resolved with \jwst\ instruments. 
On the other hand, in addition to the successful identification of the BL AGN at $z=8.50$, we identify the X-ray luminous AGN at $z=10.07$ and several potential AGN sources from multiple angles, owing to the deep NIRSpec/prism spectroscopy leveraged by the lensing effect and the ancillary deep X-ray data \citep[e.g.,][]{bogdan2023}. 
Although further observational evidence is necessary, the AGN confirmation, at least from ID10646 and ID3686, suggests the increase of the AGN fraction towards the bright end. Such a trend has been confirmed at $z\sim2$--7 \citep[e.g.,][]{sobral2018, ono2018, bowler2021, finkelstein2022a}. 
Further comprehensive AGN searches and follow-up observations may unveil even higher AGN fractions at $z\gtrsim9$ and provide the reasonable answer to the origin of the high abundance of UV-bright galaxy candidates at $z\gtrsim9$ (Section~\ref{sec:uvlf_model}).

\section{Bubbles in the Shadow at $\lowercase{z}=8.5$}
\label{sec:bubble}

\begin{figure*}
\begin{center}
\includegraphics[trim=0cm 0cm 0cm 0cm, clip, angle=0,width=1.\textwidth]{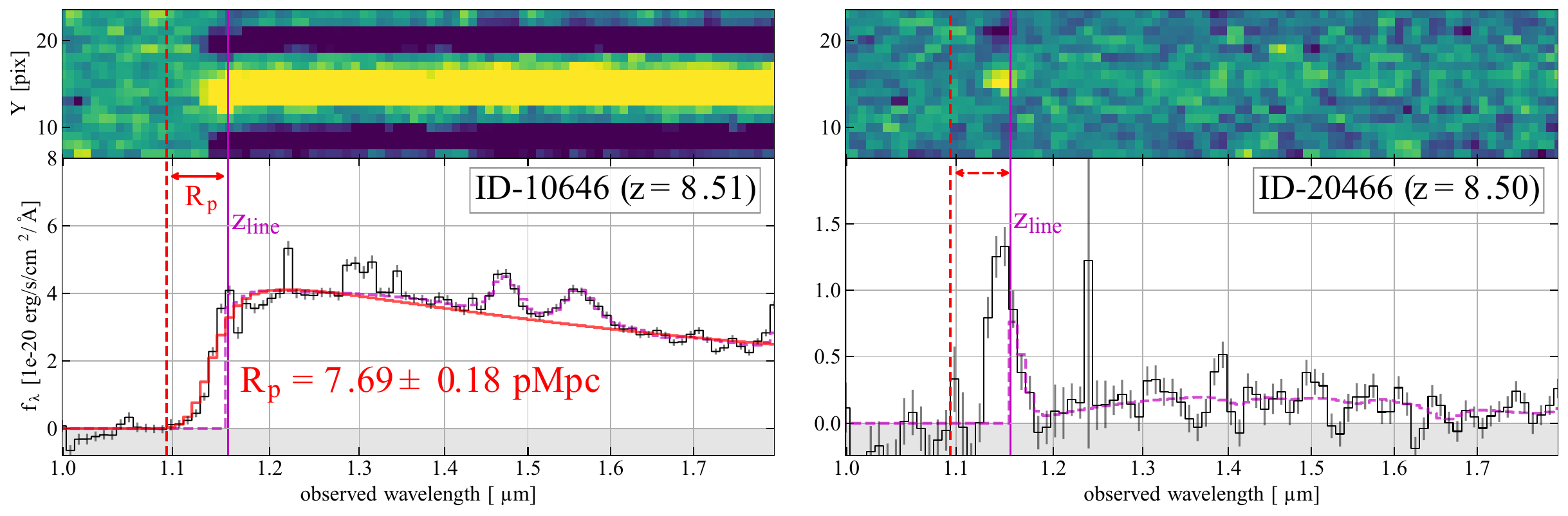}
\end{center}
\vspace{-0.6cm}
 \caption{
Zoom-in 1D+2D prism spectra of ID10646 (\textit{left}) and ID20466 (\textit{right}).
The magenta vertical line and dashed curve indicate the expected Ly$\alpha$ line wavelength and the best-fit spline model using \texttt{msaexp} based on $z_{\rm line}$, respectively. 
The blueward of the Ly$\alpha$ continuum or line is clearly detected in both sources. 
The red curve in ID10646 represents our best-fit IGM absorbed model with an ionized bubble, yielding the best-fit $R_{\rm p}$ value of $7.69\pm0.18$~pMpc. 
The red vertical line corresponds to the expected wavelength of the Ly$\alpha$ line based on the redshift of the ionization front of the ionized bubble along the line of sight.
Because of the lack of the blueward Ly$\alpha$ continuum and the difficulty in modeling the intrinsic Ly$\alpha$ line profile in ID20466, we perform the $R_{\rm p}$ measurement only for ID10466. For reference, the red dashed vertical line in ID20466 is drawn at the same wavelength as the red vertical line in the left panel. 
\label{fig:zoomin}}
\end{figure*}

\begin{figure}
\begin{center}
\includegraphics[trim=0cm 0cm 0cm 0cm, clip, angle=0,width=0.5\textwidth]{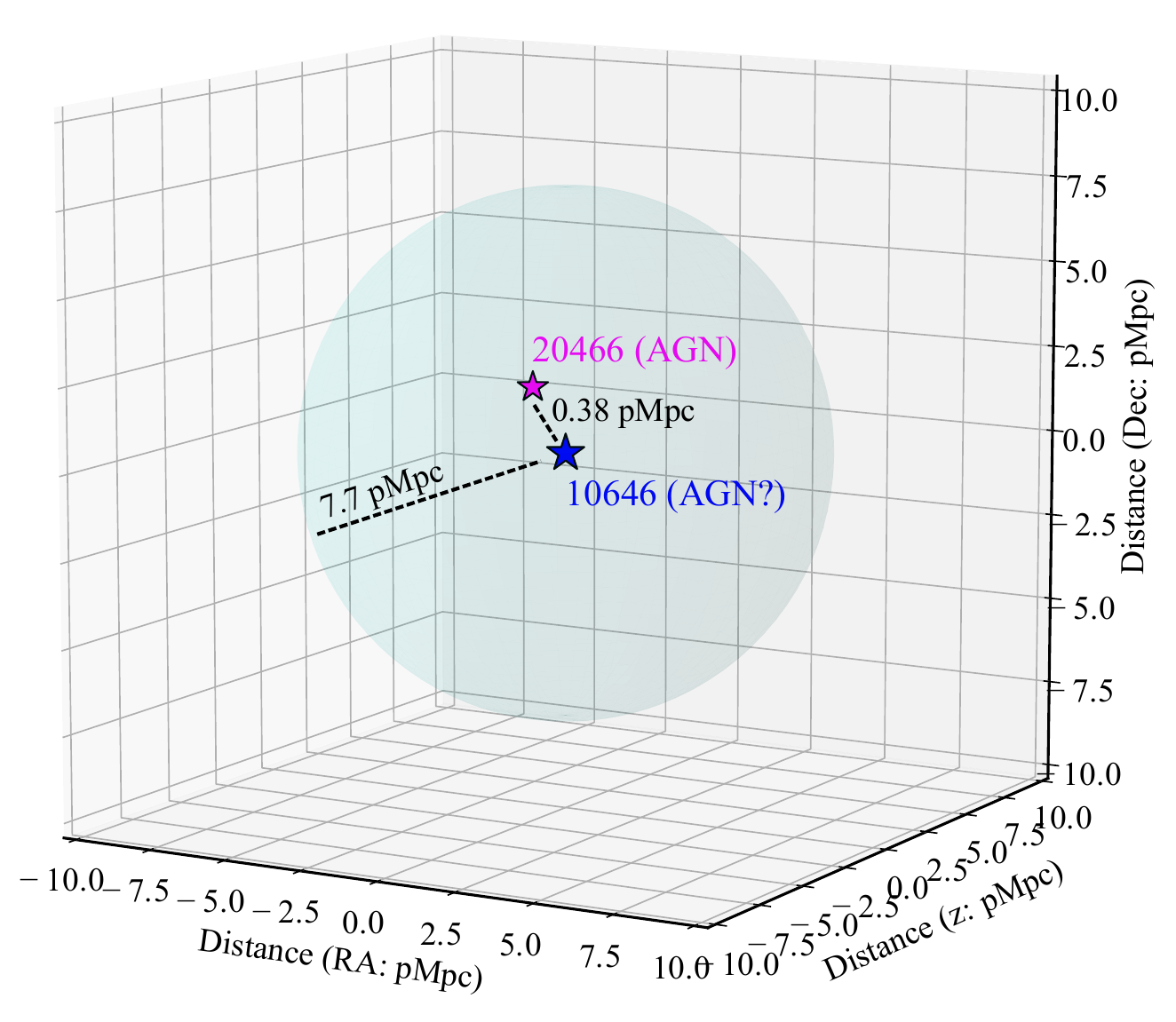}
\end{center}
\vspace{-0.2cm}
 \caption{
Schematic illustration of the relative 3D positions of ID20466 and ID10646 in the source plane. The source plane coordinates of ID20466 and ID10646 are (RA, Dec) = (3.63506157, $-30.44818861$) and (3.63209972, $-30.46295758$), respectively. 
The light blue shaded sphere denotes an isotropic sphere of the ionized bubble with $R_{\rm p}=7.7$~pMpc measured from the blueward Ly$\alpha$ line and continuum emission in the prism spectrum of ID10646. 
Their physical distance of 0.38~pMpc indicates that ID20466 also resides in the same ionized bubble, which facilitates the blueward Ly$\alpha$ emission also observed from ID20466, and the AGN activity in ID20466 (and ID10646) may contribute to forming this ionized bubble. 
\label{fig:bubble}}
\end{figure}

In Section~\ref{sec:measure}, we find that the $z_{\rm break}$ measurements are underestimated in ID20466 and ID10646, compared to their $z_{\rm line}$ measurements. This is caused by the non-zero fluxes at the blueward of the Ly$\alpha$ emission in the prism spectra. 
Such an enhanced transmission of the Ly$\alpha$ line and continuum is thought to be attributed to the presence of the so-called proximity zone, which has often been observed around high-redshift luminous quasars due to their strong radiation making the surrounding IGM neutral gas fully ionized \citep[e.g.,][]{eilers2017}. 
In addition to the unambiguous AGN feature observed in ID20466 via the broad H$\beta$ line, ID10646 also shows several uniquely high ionization lines (e.g., \niv1487, \civ1549, \heii\,1640), suggesting the presence of an AGN and/or star-forming activities that produce the strong radiation. 
Furthermore, ID20466 and ID10646 turn out to be the same redshift with a physical distance of 380~kpc in the source plane, where they might reside in the same ionized bubble. 

Figure~\ref{fig:zoomin} presents the zoom-in 2D+1D prism spectra of ID10646 and ID20466 with the expected wavelength of the Ly$\alpha$ break based on $z_{\rm line}$ (magenta vertical line).   
In both spectra, we clearly identify the emission in the blueward of the Ly$\alpha$ break, indicating the presence of ionized bubbles around these systems.  
\tcb{Note that NIRSpec meets the requirement of wavelength accuracy of 1/8 of a resolution element \citep{boker2023}, corresponding to $\sim$1/4 pixel, and thus the uncertainty of the wavelength calibration is not the cause.}
To evaluate the proximity zone radius $R_{\rm p}$, we model the IGM absorption in the same manner as \cite{totani2006}. 
For the intrinsic rest-frame UV spectrum, we perform a power-law fit at 1.2--2.0$\mu$m, masking the $\pm0.05\mu$m range of the bright emission lines detected in the spectrum. 
We then convolve the best-fit power law with the spectral resolution of NIRSpec/prism, multiply the IGM absorption model, and infer the best-fit $R_{\rm p}$ value from the minimum $\chi^{2}$ method. 
Although the blueward of the Ly$\alpha$ line is also clearly detected from ID20466, we conduct this measurement only for ID10646, 
because of the difficulties from i) the absence of the blueward of the Ly$\alpha$ continuum in the spectrum, ii) the uncertainty in modeling the intrinsic SED with its heavily dust-reddened nature, and iii) the uncertainty in modeling the intrinsic Ly$\alpha$ line profile with the current spectral resolution.

In Figure~\ref{fig:zoomin}, we show the best-fit IGM absorbed SED for ID10646 (red curve). We obtain the best-fit value of $R_{\rm p}=7.69\pm0.18$ proper Mpc (pMpc). 
To evaluate the potential effect from a weak Ly$\alpha$ line whose flux may spread into a Gaussian on both sides of the Ly$\alpha$ break \citep{jones2023}, we also test the fit, including a Gaussian component on the power law. However, the measured $R_{\rm p}$ values consistently remain at $R_{\rm p}>7$~pMpc with the rest-frame Ly$\alpha$ equivalent widths of 10--100${\rm \AA}$. 
We thus use the best-fit value from the single power-law fit as a fiducial estimate in this paper. 
In Figure~\ref{fig:bubble}, we also illustrate the ionized bubble and relative positions of ID10646 and ID20466 in the source plane. 
The best-fit $R_{\rm p}$ value exceeds the physical distance between ID20466 and ID10646. This indicates that these two sources reside in the same ionized bubble. 
Since ID20466 is the BL AGN, the ongoing and/or past quasar active phases in ID20466 might play a key role in contributing to forming the ionized bubble.
Yet, we cannot rule out the possibility that ID10646 is the main driver of the formation of this ionized bubble, given its numerous detections of the high ionization lines. 
A galaxy overdensity might also be related, where many faint galaxies contribute to forming the ionized bubble. 

\begin{figure*}
\begin{center}
\includegraphics[trim=0cm 0cm 0cm 0cm, clip, angle=0,width=0.75\textwidth]{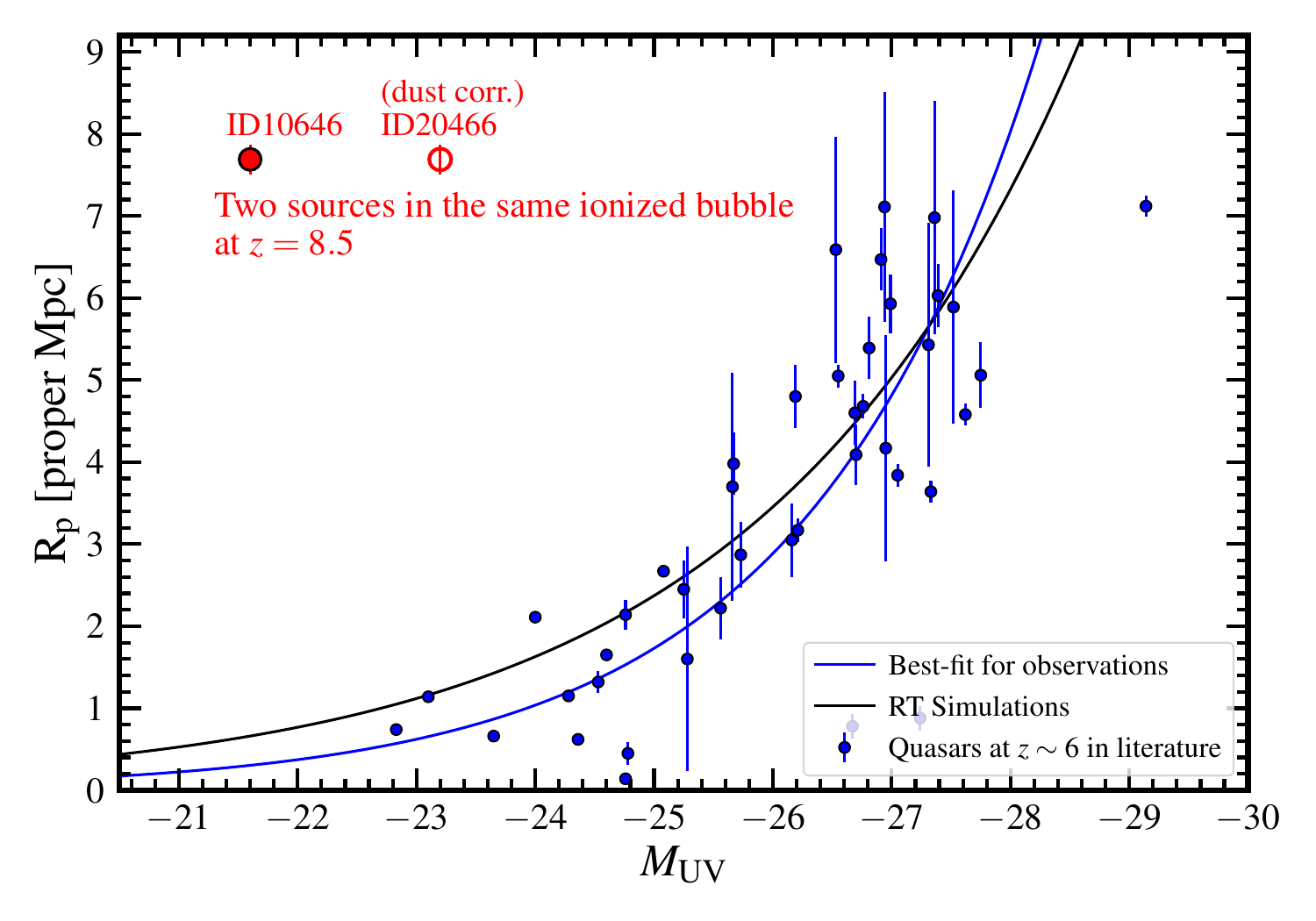}
\end{center}
\vspace{-0.2cm}
 \caption{
Sizes of the proximity zones ($R_{\rm p}$) as a function of $M_{\rm UV}$. 
The red-filled circle indicates the potential AGN source of ID10646, and the red open circle represents the dusty BL AGN of ID20466 after the dust correction. 
The blue circles indicate the previous measurements for luminous quasars at $z\sim6$ \citep{eilers2018, ishimoto2020}. 
The blue and black curves are the best-fit relations for the observation \citep{ishimoto2020} and the radiative transfer simulations \citep{eilers2017}.
ID10646 and ID20466 show remarkable gaps compared to the previous measurements and theoretical trends. 
In line with the bright Ly$\alpha$ line detection from the dusty BL AGN of ID20466, 
one interpretation is that the IGM gas density around these two sources is lower than the environment around the luminous quasars at $z\sim6$ and illuminated by the ionizing (and Ly$\alpha$) photons escaped from the dusty cloud with a low covering fraction around the AGN.  
Together with the high AGN abundance from the UVLF measurements at $z\sim9$--10 (Section~\ref{sec:uvlf}),
the identification of one of the most giant ionized bubbles to date around the dusty BL AGN and the possible AGN source indicates the non-negligible contributions of AGNs to cosmic reionization.  
\label{fig:muv-rp}}
\end{figure*}

In Figure~\ref{fig:muv-rp}, we also compare our $R_{\rm p}$ measurement with those of luminous quasars at $z\gtrsim6$ in the literature \citep[][]{eilers2017, ishimoto2020}. Because ID20466 also resides in the same ionized bubble and might play a key role in its formation, we also show the dust-corrected measurement for ID20466.\footnote{We use the bolometric luminosity estimate of $L_{\rm bol}=6.6\times10^{45}$~erg~s$^{-1}$ \citep{kokorev2023b} and assume the bolometric luminosity correction factor of 4.5 at 1500${\rm \AA}$ \citep{richards2006}. }
For comparison, we also present the best-fit $R_{\rm p}-M_{\rm UV}$ relation for the $z\sim6$ quasars estimated in \cite{ishimoto2020} and the trend obtained from the radiative transfer simulations presented in \cite{eilers2017}. 
These empirical and theoretical predictions show the positive correlation between $M_{\rm UV}$ and $R{\rm p}$. 
Remarkably, we find that ID10646 and ID20466 fall in significantly faint $M_{\rm UV}$ and large $R_{\rm p}$ parameter space, $\sim1$--2 orders of magnitude fainter $M_{\rm UV}$ than what is predicted from the typical positive correlation obtained both from the previous observational and theoretical results. 
These remarkable gaps indicate the following three possible interpretations: i) ID20466 and/or ID10646 had more luminous quasar phases until recently, ii) the IGM gas density is much lower around these two sources than those of the luminous quasars at $z\sim6$, which enables to form a large $R_{\rm p}$ with a relatively low luminosity\footnote{Str\"{o}mgren radius $\propto L^{1/3} \times n_{\rm gas}^{-2/3}$}, and iii) there exist many faint galaxies around that are the main drivers of the ionized bubble.   
If the first interpretation is the case, the correlations of $M_{\rm UV}$ and $R_{\rm p}$ suggest that the more luminous phase of these objects could reach $M_{\rm UV}\lesssim-26$. 
From the survey volume in UNCOVER at $z\sim8.5$, the predicted number of such luminous quasars ($M_{\rm UV}\lesssim-26$) is $\ll 10^{-5}$ \citep[e.g.,][]{dayal2019}, indicating that the presence of such a single luminous quasar is already extremely challenging to our current BH formation and evolution models.  
In the interpretation of ii), given that Str\"{o}mgren radius $\propto n_{\rm gas}^{-2/3}$, a gas density lower than the environment around the luminous quasars by a factor of $\sim100$ may explain that ID20466 provides a sufficient amount of the ionizing photon budget to form the observed proximity zone. 
Nevertheless, this interpretation still requires the covering fraction of the dusty cloud around the AGN to be low, where most ionizing photons successfully escape from the system to ionize the surrounding neutral IGM. 
In the interpretation of iii), although we do not find evidence of a galaxy overdensity at $z\sim8.5$ in our SED catalog \citep{bwang2024a}, these galaxies fall close to the edge of the primary area of the UNCOVER NIRCam observations (Figure~\ref{fig:entire}), which might make it difficult to identify the nearby faint galaxies at the same redshift.

Interestingly, we also observe the Ly$\alpha$ line from ID20466, despite its heavily dusty nature with $A_{\rm V}=2.1$ (Section~\ref{sec:measure}). 
The giant ionized bubble may facilitate the Ly$\alpha$ line, including the blueward emission, while the heavily dusty nature indicates that the origin of the Ly$\alpha$ line is not the emission of the BL AGN from the line of sight. Instead, there are the following three possible scenarios for the origin of the Ly$\alpha$ line from ID20466: 
a) cold gas inflow, 
b) a blue, un-obscured host galaxy, and 
c) scattered light escaped from angles different from the dust-obscured AGN line of sight. 

In the scenario of a), we include a single Gaussian to the power law function, convolve with the wavelength-dependent spectral resolution of the prism, and fit it to the prism spectra at $0.9\text{--}1.6\,\mu\mathrm{m}$ to measure the velocity offset of Ly$\alpha$ line peak. We find that the Ly$\alpha$ peak is blue-shifted by $2500\pm300\,\mathrm{km}\,\mathrm{s}^{-1}$ ($\simeq0.01\,\mu\mathrm{m}$) with respect to the systemic redshift determined by other emission lines ($z_{\rm line}$). However, the velocity scale of the cold gas inflow is regulated by the gravitational potential \cite[e.g.,][]{larson2019}, which would be comparable to the escape velocity of $\sqrt{2GM_{\rm h}/R_{\rm vir}}$, where $M_{\rm h}$ and $R_{\rm vir}$ are the halo mass and virial radius, respectively. Assuming the maximum $M_{\rm h}$ of $\sim10^{11}\,M_{\rm h}$ at $z=8.5$ from our survey volume \cite{behroozi2020} and its $R_{\rm vir}$, the dynamical velocity is estimated to be $\sim200\,\mathrm{km}\,\mathrm{s}^{-1}$. While the resonance scattering nature of the Ly$\alpha$ line can increase the velocity offset by a factor of $\sim2$ \cite[e.g.,][]{verhamme2006}, this indicates that the measured blue-shifted velocity is still too high to be explained by the cold gas inflow.  
Instead, a degeneracy between velocity and morphology arises if the Ly$\alpha$ emission is spatially extended and the emitting peak position differs from other emission lines within the MSA shutter. 
Based on the line spread function for a uniformly-illuminated slit and assuming the resolution element for the prism is 2.2 pixels \citep{jakobsen2022}, we estimate this effect to be up to $\sim0.016$~$\mu$m at 1.1~$\mu$m if the peak of the Ly$\alpha$ distribution is located on the opposite side of the microshutter compared to the other line-emitting gas. The measured Ly$\alpha$ offset is equal to $\sim0.01$~$\mu$m, indicating that the Ly$\alpha$ offset in the prism may be explained by this differential morphology effect within the shutter, although we do not find any clear evidence of the differential morphology in the F115W that includes the Ly$\alpha$ line emission. 
To give a definitive answer, we need a high spectral resolution follow-up and the detailed Ly$\alpha$ line profile. 
In the scenario of b), it might be the case that the AGN core is heavily dust reddened, while the host galaxy is a blue, un-obscured galaxy from which Ly$\alpha$ is emitted. From the Gaussian and power-law fitting above, we obtain the Ly$\alpha$ flux of $\simeq3.7\times10^{42}$~erg~s$^{-1}$. Assuming no dust attenuation and the IGM absorption, we estimate SFR from this Ly$\alpha$ luminosity of $\simeq$4~$M_{\odot}$/yr using a calibration of \cite{kennicutt1998}. This SFR value is comparable to a UV luminosity observed at 27.8~mag in F150W, using the \cite{kennicutt1998} calibration. However, ID20466 is as faint as 28.8~mag in F150W with the compact morphology, indicating that the emission in F150W is still dominated by the AGN component and that the UV magnitude of the host galaxy is $\gg28.8$~mag. 
\tcb{Although there is a scatter between the H$\beta$/UV luminosity ratio \citep[e.g.,][]{simmonds2023},
these properties might indicate that the scenario of b) is unlikely.} 
In the scenario of c), the strong Ly$\alpha$ emission is originally from the AGN, and its resonance scattering nature helps to avoid the dusty cloud in the line of sight.  
It has also been argued that the rest-UV light in these red compact objects like ID20466 may be caused by the scattered light \citep[e.g.,][]{assef2020, glikman2023, noboriguchi2023} based on their unique SED shape of the blue in UV and red in optical colors \citep{labbe2023}. 
\tcb{
In this process, the Ly$\alpha$ may be scattered and thus avoid the dusty clouds, similarly to the rest-UV light, while the ionizing photons can escape to the IGM from the holes of neutral hydrogen and may directly contribute to forming the ionized bubble around ID20466.
It is worth mentioning that the scattered Ly$\alpha$ line may be extended, so-called Ly$\alpha$ halo \citep[e.g.,][]{leclercq2020}, which is also well aligned with the ``blue-shifted'' Ly$\alpha$ peak observed in ID20466, because the blue-shifted Ly$\alpha$ peak is explained by the differential morphology effect within shutter, as described above.  
}

In short, c) is the most likely scenario among these three, 
which is also in line with the interpretation of the remarkable gap observed in the $M_{\rm UV}-R_{\rm p}$ relation that ID20466 (and ID10646) forms the giant ionized bubble in a relatively low-density IGM environment (interpretation ii).  
Despite the small survey volumes, the dozens of dust-reddened compact objects have been identified in recent \hst\ and \jwst\ studies at $z\sim4-7$ \citep[e.g.,][]{endsley2023, fujimoto2022, furtak2023, labbe2023}, and some have already been spectroscopically confirmed to be AGNs from the BL Balmer line detection \citep{kocevski2023, matthee2023, furtak2023c}. 
These identifications indicate a high abundance of the dusty AGNs at high redshifts, resulting in steeper faint-end slopes of AGN LFs at $z\sim4-7$ \citep[e.g.,][]{labbe2023, matthee2023} than what is estimated from previous Type-I quasar measurements \citep[e.g.,][]{akiyama2018, mcgreer2018, matsuoka2018}. While their dusty nature has implied that their contributions to cosmic reionization are minimal, the discovery of the giant ionized bubble and the bright Ly$\alpha$ line detected from ID20466 suggests its potential contributions to forming the ionized bubble, despite its heavily dusty nature. 
Importantly, a similarly bright Ly$\alpha$ line is also observed in another dusty BL AGN at $z=7.0$ ($A_{\rm V}\sim3$; \citealt{furtak2023c}). If a situation similar to ID20466 is also taking place in other dusty BL AGNs, the ionizing photon escape from dusty AGNs may be a recurrent event at the heart of the epoch of cosmic reionization. While recent spectroscopic observations for faint galaxies (down to $M_{\rm UV}=-15$) at $z\simeq6$--8 provide firm evidence that the main driver of cosmic reionization is galaxies \citep{atek2023b}, our results imply that the abundant (dusty) AGNs at $z\gtrsim9$ may still have non-negligible contributions to cosmic reionization.

\section{Summary}
\label{sec:summary}
In this paper, we present \jwst\ NIRSpec prism follow-up observations in the multi-object spectroscopy mode (MOS) using the multi-shutter array (MSA) for $z\gtrsim9$ galaxy candidates.  These candidates were identified in the Cycle~1 Treasury program of UNCOVER (\#2561, PIs: I.~Labbe \& R.Bezanson; \citealt{bezanson2022}) behind the massive galaxy cluster A2744.
Owing to its extensive designs of the NIRSpec observations among the lensing cluster surveys in \jwst\ Cycle~1, these new deep prism spectra, leveraged by the gravitational lensing effect, afford us unparalleled opportunities to perform the initial spectroscopic census of these early galaxies and investigate the UV luminosity function (LF), the AGN fraction, and their contributions to cosmic reionization out to the redshift frontier in a wide UV luminosity range. 
The major findings of this paper are summarized below: 
\begin{enumerate}
\item Of the 680 distinct targets in our NIRSpec MSA, we confirm the source redshift via emission lines and/or the Ly$\alpha$ break feature in the prism spectra for 10 lensed galaxies ($\mu=$1.3--12.8) ranging from $z=8.50$ to $13.08$. These galaxies have a $M_{\rm UV}$ range of $\in [-21.72:-17.28]$ after lensing corrections. This increases the spec-$z$ confirmed sample so far known in a high-redshift ($z\gtrsim9$) and UV-faint ($M_{\rm UV}\gtrsim-19$) regime by a factor of 3.
\item Although the ten spec-$z$ confirmed sources are initially selected through several different selection criteria, four sources are the sample selected from a systematic NIRCam analysis presented in \cite{atek2023} (hereafter A23), which achieves a high confirmation rate of 100\%. 
Six sources show robust multiple emission line detections, providing the most secure redshift estimates. 
The other four sources are mainly constrained with the Ly$\alpha$ break feature. 
\item For the homogeneous sample from A23, we do not find systematic overestimates in $z_{\rm phot}$ from $z_{\rm spec}$ reported in recent NIRSpec studies, probably owing to the deep blue NIRCam filters (F115W, F150W) taken in UNCOVER and some strict selection criteria adopted in the A23 selection.  
Using six sources with multiple emission line detection, we also evaluate the offset of the redshift estimates between the lines ($z_{\rm line}$) and the Ly$\alpha$ break ($z_{\rm break}$). We find that the offset can be as large as out to $\pm0.2$. This offset can be even worse with lower S/N data, which raises caution in designing future follow-up spectroscopy for the break-only sources, especially with ALMA. 
\item In addition to the X-ray luminous AGN confirmed at $z=10.07$ \citep{goulding2023}, we newly identify a dusty broad-line (BL) AGN at $z=8.50$ \citep{kokorev2023b}. Besides, the prism spectra for the two most UV-luminous galaxies in our spec-$z$ sample cohort hint at AGN activity. This is inferred from several highly-ionized gas emission lines detected at high significance levels (e.g., \niv1487, \civ1549, \heii1640) and an elevated \oiii5008/H$\beta$ ratio exceeding 10 observed in ID10646 and ID3686. 
\item In conjunction with the spec-$z$ confirmed sources in UNCOVER and other general field surveys of CEERS and JADES, we infer lower bounds on the UV LFs at $z\sim9$, $z\sim10$, and $z\sim12$. Our results align with previous photometric measurements and improve the lower constraints previously established from recent spectroscopic studies.  Our results also confirm the high abundance of the UV-bright ($M_{\rm UV}\lesssim-20$) galaxies at $z\gtrsim9$, which challenges several current theoretical models. In the $z\sim9$ UVLF, we find a significant excess in the brightest $M_{\rm UV}$ bin, spanning $[-22:-21]$, comprising ID10646 and ID3686. Given their different redshifts, this excess is attributable not to an overdensity but to their uniquely UV-bright properties. This further reinforces the hypothesis that these UV-luminous sources, characterized by multiple high-ionization emission lines, are indeed AGNs. 
\item With the spec-$z$ confirmed BL AGN and X-ray luminous AGN, we also evaluate the lower limits on the AGN LF at $z\sim9$ and $z\sim10$. These lower limits require the AGN LFs at $z\sim$ 9--10 comparable or even higher amplitude than the X-ray AGN LF estimated at $z\sim6$. 
\item Our results suggest a relatively high AGN fraction of $> 10$--35\% even at $z\gtrsim9$, compared to the previous reports of $\approx$~5-10\% from the BL AGN identification at $z\sim4$-7. This high AGN fraction is likely attributed to the comprehensive AGN recognition made feasible by our intensive 2.7–-11.8~hours prism exposures (cf. $\sim$1–-2.6~hours in previous NIRSpec studies for BL AGN search) and the ancillary deep X-ray data, both of which benefit substantially from the gravitational lensing effect. These results indicate the plausible cause of the high abundance of $z>9$ galaxies claimed in the recent photometric studies may be the AGNs. 
\item We identify the non-zero fluxes at the blueward of Ly$\alpha$ emission in the prism spectra of the dusty BL AGN of ID20466 at $z=8.50$ and the potential AGN source of ID10646 at $z=8.51$. The proximity zone size measurement shows that these two sources resided in the same ionized bubble with $R_{\rm p}=7.69\pm0.18$~proper Mpc. 
Both these sources notably deviate from the established $M_{\rm UV}-R_{\rm p}$ relationship observed in luminous quasars at $z\sim6$. Despite its heavily dusty nature with $A_{\rm V}=2.1$, the Ly$\alpha$ line is also detected from ID20466. A plausible explanation is that the covering fraction of the dusty cloud surrounding the AGN is minimal, thereby facilitating significant ionizing photon escape. This is in line with our identification of the giant ionized bubble. Our results, taken in concert with indications of the high AGN fraction even at $z\gtrsim9$, suggest that AGNs might have played a non-negligible role during cosmic reionization.
\end{enumerate}

We thank the anonymous referee for the helpful and constructive review on the paper. 
We also thank Peter Larson, Darach Watson, Charlotte Mason, Kohei Inayoshi, and Akio Inoue for helpful discussions on the proximity zone size measurements and the interpretation of the ionized bubble identified at $z=8.5$. 
We are also grateful to Steven Finkelstein and Pablo Arrabal-Haro for valuable discussions and for sharing the materials for the survey volume calculations for the NIRSpec programs conducted in the CEERS field.  
We also thank Rohan Naidu for discussing the spec-$z$ confirmation results for $z>10$ galaxies, Nikko Cleri for discussing the UV emission line properties, and Maximilien Franco for sharing the UVLF measurements. 
This work is based on observations made with the NASA/ESA/CSA James Webb Space Telescope. The data were obtained from the Mikulski Archive for Space Telescopes at the Space Telescope Science Institute, operated by the Association of Universities for Research in Astronomy, Inc., under NASA contract NAS5-03127 for JWST. 
These observations are associated with program JWST-GO-2561. Support for program JWST-GO-2561 was provided by NASA through a grant from the Space Telescope Science Institute, which is operated by the Association of Universities for Research in Astronomy, Inc., under NASA contract NAS 5-03127. 
This project has received funding from NASA through the NASA Hubble Fellowship grant HST-HF2-51505.001-A awarded by the Space Telescope Science Institute, which is operated by the Association of Universities for Research in Astronomy, Incorporated, under NASA contract NAS5-26555.
H.A. and IC acknowledge support from CNES, focused on the JWST mission, and the Programme National Cosmology and Galaxies (PNCG) of CNRS/INSU with INP and IN2P3, co-funded by CEA and CNES. 
PD acknowledges support from the NWO grant 016.VIDI.189.162 (``ODIN") and from the European Commission's and University of Groningen's CO-FUND Rosalind Franklin program.
AZ and LJF acknowledge support by Grant No. 2020750 from the United States-Israel Binational Science Foundation (BSF) and Grant No. 2109066 from the United States National Science Foundation (NSF), and by the Ministry of Science \& Technology, Israel. 
This work has received funding from the Swiss State Secretariat for Education, Research and Innovation (SERI) under contract number MB22.00072, as well as from the Swiss National Science Foundation (SNSF) through project grant 200020\_207349.
The Cosmic Dawn Center (DAWN) is funded by the Danish National Research Foundation under grant No.\ 140. The work of CCW is supported by NOIRLab, which is managed by the Association of Universities for Research in Astronomy (AURA) under a cooperative agreement with the National Science Foundation. 
MS acknowledges support from the CIDEGENT/2021/059 grant, from project PID2019-109592GB-I00/AEI/10.13039/501100011033 from the Spanish Ministerio de Ciencia e Innovaci\'on - Agencia Estatal de Investigaci\'on. MST also acknowledges the financial support from the MCIN with funding from the European Union NextGenerationEU and Generalitat Valenciana in the call Programa de Planes Complementarios de I+D+i (PRTR 2022) Project (VAL-JPAS), reference ASFAE/2022/025.

Some/all of the data presented in this paper were obtained from the Mikulski Archive for Space Telescopes (MAST) at the Space Telescope Science Institute. 
The specific observations analyzed can be accessed via \dataset[10.17909/bvf5-mp20]{http://dx.doi.org/10.17909/bvf5-mp20}. 
The reduced data is also available via \url{https://jwst-uncover.github.io/}.

\software{\texttt{msaexp} (0.6.10; \citealt{brammer2022b}),  \texttt{astropy} \citep{astropy2013, astropy2018, astropy2022}, \texttt{eazy} \citep{brammer2008}, \texttt{FSPS} \citep{conroy2010}, Prospector \citep{johnson2021}. 
}

\clearpage
\appendix

\section{Details of Individual Sources}
\label{sec:app_ind}

Below, we briefly summarize the information for the spec-$z$ confirmed ten UNCOVER sources presented in this paper. 

{\flushleft \bf \boldmath 20466 ($z=8.50$)} -- 
This source is included in the MSA as a high-redshift dusty AGN candidate due to the red color and compact morphology (ID13556 in \citealt{labbe2023}). 
We observed this source once in MSA2 with an exposure time of 2.7 hours.  
In addition to the Ly$\alpha$ break feature, multiple emission lines are detected at SNR $\geq$ 2.5, such as Ly$\alpha$, Mg\,{\sc ii}\,$\lambda\lambda$2796, 2803, \neiii3869, \neiii3968, \oiii4363, \oiii4960, \oiii5008, H$\delta$, H$\gamma$, and H$\beta$, constraining its redshift at $z=8.500^{+0.000}_{-0.001}$. 
The unambiguous broad line (BL) emission is identified in H$\beta$, being the highest redshift BL AGN whose Balmer BL is securely (SNR $>$ 5) detected so far \citep[see also][]{larson2023}. 
The Ly$\alpha$ line, bluer than its rest-frame 1216${\rm \AA}$, is also detected, and we discuss its possible physical origins in Section~\ref{sec:bubble}. 
The photometric catalog ID is 21347 \citep{weaver2023}. 
Further details of the sample selection and characterizations will be presented in J.~Greene et al. in prep. and \cite{kokorev2023b}.

{\flushleft \bf \boldmath 10646  ($z=8.51$)} -- 
This source is included in the MSA due to its uniquely red color (F277W--F444W = 1.2~mag). 
We observed this source once in MSA2 with an exposure time of 2.7 hours. 
In addition to the unambiguous Ly$\alpha$ break, multiple emission liens are detected at SNR $\geq2.5$, such as \niv1487, \civ1549, \oiiib1661,1666, \heii1640, \ciii1907,1909, Mg\,{\sc ii}\,$\lambda\lambda$2796,2803, \nev3346, \oii3727,3730, \neiii3869, \neiii3968, \oiii4363, \oiii4960, \oiii5008, H$\gamma$, H$\delta$, H$\beta$, and He\,{\sc i}, constraining its redshift at $8.511^{+0.000}_{-0.001}$. 
Note that the \nev3426/\nev3346 ratio is almost constant at 2.73. We confirm \nev3426 is also observed in our spectrum with SNR $\sim2$, while it indicates \nev3426/\nev3346 $\sim1$. This might suggest that the \nev3346 line detection could be spurious, although it is challenging to conclude with their current SNRs. 
ID10646 is spatially separated from ID20466 by a physical scale of 380~kpc in the source plane, and their redshift difference is only 0.01. We thus interpret these sources as residing in the same massive dark matter halo. 
Similar to ID20466, the continuum blueward of the Ly$\alpha$ line is also detected in ID10646, and we discuss its possible physical origins in Section~\ref{sec:bubble}.  
ID10646 is uniquely UV bright. With $M_{\rm UV}=-21.5$~mag, it is comparably bright to GNz11 \citep[e.g.,][]{bunker2023} and shows several highly ionized gas emission lines at high significance levels (e.g., \niv1487, \civ1549, \heii1640). However, we cannot rule out the possibility that the galaxy emission is driven by star-forming activity rather than AGN  based on rest-frame UV-optical line diagnostics alone \citep[e.g.,][]{feltre2016}. 
The photometric catalog ID is 11701 \citep{weaver2023}.
Further details and characterizations will be presented in J.~Weaver et al. in prep.

{\flushleft \bf 3686 \boldmath ($z=9.33$)} -- 
This source is included in the MSA as one of the robust $z>9$ candidates selected in \cite{atek2023} (hereafter A23). 
We observed this source once in MSA2 with an exposure time of 2.7 hours. 
In addition to the unambiguous Ly$\alpha$ break, multiple emission lines are detected at SNR $\geq$ 2.5, such as \niv1487, \nev3426, \oii3723,3730, \neiii3869, \oiii4960, \oiii5008, and H$\gamma$, constraining its redshift at $z=9.325_{-0.001}^{+0.000}$. 
As part of the GLASS-\jwst\ survey (\#1324; PI. T.~Treu; \citealt{treu2022}) and a follow-up DDT program (\#2756; PI. W.~Chen), a consistent NIRSpec/prism spectroscopic confirmation has also been reported in \cite{boyett2023}, while the previous prism observations only cover the wavelength range of $\sim1.1$--4.5~$\mu$m with detector gaps.  
The full prism spectrum coverage of $\sim0.6$--5.2$\mu$m newly detects several emission lines from this source, including \niv1487, H$\beta$ and \oiii4960, 5008. 
On the other hand, the \nev3426 line is not detected in the previous observations, although its observed wavelength was covered by the previous observations. This suggests that the \nev3426 line could be spurious, while the shutter configurations of MSA are not exactly the same between previous and our observations. 
The source is spatially extended, indicative of an interacting system \citep{boyett2023}. 
ID3686 is the most luminous high-redshift galaxy candidate at $z\gtrsim9$ in the original photometric catalog with $M_{\rm UV}=-21.7$. 
Employing a nearby empty shutter, we also produce spectra for the three shutters with the global background subtraction, where the \oiii5008/H$\beta$ shows uniquely high ratios of $\sim11$--18 in the central and Southern East shutters (see Section~\ref{sec:measure}). These ratios exceed the maximum value of $\sim10$ observed in recent NIRSpec studies for galaxies at $z\sim2$-9 and fall in the AGN regime in the \nii, [S\,{\sc ii}], and [O\,{\sc i}] BPT diagrams \citep{sanders2023}. 
Such a high ratio may also be induced by the shock excitation \citep[e.g.,][]{kewley2013, hirschmann2022}, while the fact that the similarly high ratio also observed in the Southern East shutter, where the emission is dominated by the compact component in the NIRCam map, indicates that these line properties may be caused by a strong radiation of an AGN. 
The photometric catalog ID is 4745 \citep{weaver2023}.

{\flushleft \bf 22223 \boldmath ($z=9.57$)} -- 
This source is included in the MSA as one of $z>9$ candidates selected from the SED analysis using \texttt{eazy} and \texttt{prospector} \citep{bwang2024a}. 
We observed this source once in MSA4 with an exposure time of 4.4 hours. 
In addition to the unambiguous Ly$\alpha$ break, multiple emission lines are detected at SNR $\geq$ 2.5, including \civ1549, \oiii4960, \oiii5008, H$\delta$, H$\gamma$, and H$\beta$. 
In the 2D spectrum, we also identify an unknown line at $\sim$0.9$\mu$m, probably due to a failed open shutter, though our extraction does not include either the positive and negative features from this line, and thus this does not affect our results.   
The prism spectrum shows a softened Ly$\alpha$ break shape, also reported in other $z>9$ prism-observed galaxies. This shape is likely caused by some combination of effects of the Ly$\alpha$ damping wing, the intrinsic SED shape, and/or an additional DLA system \citep[e.g.,][]{curtis-lake2023, arrabal-haro2023b, umeda2023, heintz2023}. 
The photometric catalog ID is 23089 \citep{weaver2023}.

{\flushleft \bf 31028 \boldmath ($z=9.74$)} -- 
This source is included in the MSA as one of $z>9$ candidates with a high magnification ($\mu>5$) selected from SED analysis using \texttt{eazy} and \texttt{prospector} \citep{bwang2024a}. 
We observe this source in MSA3 and MSA6, with a total exposure time of 6.9 hours. 
However, due to the lack of an obvious continuum trace in the 2D spectrum taken in MSA3, probably because of more significant slitloss (see Fig.~\ref{fig:cutout}) and the potential systematic uncertainty in the slitloss correction in the coadd process, we only use the data taken in MSA6 in this paper (4.4h). 
Our template fitting supports the high-$z$ solution from the Ly$\alpha$ break feature (see also the blue curve representing the forced low-$z$ best-fit solution in Fig.~\ref{fig:spectrum}), constraining its redshift at $z=9.740^{+0.000}_{-0.001}$, 
where the Ly$\alpha$ line is tentatively detected at SNR$\sim$3. 
No emission lines are detected above SNR $\geq2.5$. 
The redshift estimate subsequently leads to a magnification estimate of $\mu = 6.73^{+1.50}_{-0.05}$ and $M_{\rm UV}=-17.31$~mag, making it intrinsically the faintest source among the spec-$z$ confirmed objects at $z\geq8.5$ with \jwst\ so far \citep[e.g.,][]{williams2022, roberts-borsani2022b, fujimoto2023a, arrabal-halo2023a, arrabal-haro2023b, hainline2023}. 
The photometric catalog ID is 31955 \citep{weaver2023}.

{\flushleft \bf 13151 \boldmath ($z=9.88$)} -- 
This source is included in the MSA as one of $z>9$ candidates with a high magnification ($\mu>5$) selected from the SED analysis using \texttt{eazy} and \texttt{prospector} \citep{bwang2024a}. 
We observed this source three times in MSA5, MSA6, and MSA7, with a total exposure time of 11.8 hours. 
 In addition to the unambiguous Ly$\alpha$ break, multiple emission lines are detected at SNR $\geq$ 2.5, including \oiiib1661,1666, and \ciii1907,1909. 
 Previous NIRSpec/prism observations detect the Ly$\alpha$ break feature, providing the redshift solution of $z=9.79$ via a similar \texttt{eazy} template fitting method \cite{roberts-borsani2023}, while the multiple emission line identification and the better sensitivity in the Ly$\alpha$ break make the redshift solution firmly improved. 
The photometric catalog ID is 14088 \citep{weaver2023}.

{\flushleft \bf 26185 \boldmath ($z=10.07$)} -- 
This source is included in the MSA as one of the robust $z>9$ candidates selected in A23 (see also e.g., \citealt{castellano2023}). Moreover, an X-ray luminous AGN has been reported from a 1.25~Ms deep {\it Chandra} observation, making this the highest-$z$ X-ray AGN known \citep{bogdan2023}.  
We observe this source twice in MSA1 and MSA4, with a total exposure time of 7.1~hours. 
In addition to the unambiguous Ly$\alpha$ break, multiple emission lines are detected at SNR $\geq$ 2.5, such as \ciii1709,1909, \oii3727, 3730, \neiii3869, \neiii3968, and H$\gamma$, constraining its redshift at $z=10.071_{-0.001}^{+0.000}$. 
The photometric catalog ID is 27025 \citep{weaver2023}.
Further details and characterizations have been presented in \cite{goulding2023} as UHZ1.

{\flushleft \bf 37126 \boldmath ($z=10.23$)} -- 
This source is included in the MSA as one of the robust $z>9$ candidates selected in A23. 
We observed this source twice in MSA3 and MSA4, with a total exposure time of 6.9 hours. The source failed to be successfully extracted from MSA3 in our early reduction, and thus we use the data from MSA4 (4.4h) in this analysis.  
The peaky feature at $\sim4.8\mu$m in the spectrum is an artifact, and we mask the relevant pixels in our template fitting. From the unambiguous Ly$\alpha$ break, the redshift is securely estimated at $z=10.255^{+0.001}_{-0.001}$ with N\,{\sc iii}]\,$\lambda\lambda$1747,1749 detection at SNR $\sim$ 3. 
The photometric catalog ID is 38095 \citep{weaver2023}.

{\flushleft \bf 38766 \boldmath ($z=12.39$)} -- 
This source is included in the MSA as one of the robust $z>12$ candidates selected in A23. 
We observed this source once in MSA4, with an exposure time of 4.4 hours. 
The unambiguous Ly$\alpha$ break, a tentative He\,{\sc i} (SNR$\sim$2.5) are detected, constraining the source redshift at $z=12.393^{+0.004}_{-0.001}$. 
At a consistent redshift, tentative \oii3727,3730 and Mg\,{\sc ii}\,$\lambda\lambda$2796,2803 are also detected (SNR$\sim$2).  
Within a $\sim$2~arcmin on the sky, a remarkably UV bright galaxy is identified with a very close photometric redshift ($z_{\rm phot}=12.4^{+0.1}_{-0.3}$; \citealt{naidu2022c}, see also \citealt{castellano2022, donnan2022, harikane2023, bouwens2022d}), where there might exist a galaxy overdensity. 
The photometric catalog ID is 39753 \citep{weaver2023}.
Further details and characterizations have been presented in \cite{bwang2023} as UNCOVER-z12. 

{\flushleft \bf 13077 \boldmath ($z=13.08$)} -- 
This source is included in the MSA as one of $z>12$ candidates selected from the SED analysis using \texttt{eazy} and \texttt{prospector} \citep{bwang2024a}. 
We observe this source twice in MSA5 and MSA7, with a total exposure time of 7.4 hours. 
No emission lines are detected above SNR $\geq2.5$, while our template fitting shows the high-$z$ solution from the Ly$\alpha$ break feature (see also the blue curve representing the forced low-$z$ best-fit solution in Fig.~\ref{fig:spectrum}), constraining its redshift at $z=13.079^{+0.014}_{-0.001}$. 
Although the Ly$\alpha$ break feature is less secure than other sources, the template fittings to both individual spectra taken in MSA5 and MSA7 show the high-$z$ solution as well. 
The photometric catalog ID is 14019 \citep{weaver2023}.
Further details and characterizations have been presented in \cite{bwang2023} as UNCOVER-z13.

\section{ID\lowercase{s} in different literature}
\label{sec:app_ids}

Our ten spec-$z$ confirmed sources have also been reported in previous studies in various contexts. Here, we summarize the IDs of the ten spec-$z$ confirmed sources presented in different literature. 

\setlength{\tabcolsep}{2pt}
\begin{deluxetable}{lccc}
\tablecaption{IDs of the ten spec-$z$ confirmed sources at $z\geq8.5$ in our UNCOVER/NIRSpec observations}
\tablehead{\colhead{ID (This)}  & \colhead{ID (We23)}  & \colhead{ID (Other)} \\
        (1)   &    (2)          &          (3)  }
\startdata
20466 & 21347  & 13556 (L23) \\
10646 & 11701 &  \nodata \\
3686  & 4745 & Gz9p3 (B23), 2065 (A23), DHZ1 (C23) \\
22223 & 23089 & \nodata  \\
31028 & 31955 & \nodata  \\
13151 & 14088 & JD1 (Z14), (RB23)  \\
26185 & 27025 & UHZ1 (C23), 21623 (A23) \\
37126 & 38095 & 39704 (A23)  \\
38766 & 39753 & 42329 (A23), UNCOVER-z12 (Wa23)  \\
13077 & 14019 & UNCOVER-z13 (Wa23)  \\
\enddata
\tablecomments{
(1) Source ID used in the MSA design and this paper. 
(2) Source ID used in the UNCOVER photometric catalog of \citep{weaver2023}. 
(3) Source ID or name used in other literature (L23; \citealt{labbe2023}, A23; \citealt{atek2023}, B23; \citealt{boyett2023}, C23; \citealt{castellano2023}, RB23; \citealt{roberts-borsani2023}, G23; \citealt{goulding2023}, Wa23; \citealt{bwang2023}, Z14; \citealt{zitrin2014}). 
}
\label{tab:id}
\end{deluxetable}

\bibliographystyle{apj}
\bibliography{apj-jour,reference}

\end{document}